\documentclass[12pt,preprint]{aastex}

\pdfoutput=1

\usepackage{epsfig}
\usepackage{natbib}
\def\plotone#1{\centering \leavevmode
\includegraphics[clip=, width=.85\columnwidth]{#1}}
\def\plottwo#1#2{\centering \leavevmode
\includegraphics[width=.45\columnwidth]{#1} \hfil
\includegraphics[width=.45\columnwidth]{#2}}

\newcommand{\cN}[1]{\mathcal{N}}

\def\gsim{\;\rlap{\lower 2.5pt
 \hbox{$\sim$}}\raise 1.5pt\hbox{$>$}\;}
\def\lsim{\;\rlap{\lower 2.5pt
   \hbox{$\sim$}}\raise 1.5pt\hbox{$<$}\;}

\begin{document}


\title{Habitable Climates: The Influence of Obliquity}

\author{David S. Spiegel\altaffilmark{1,2}, Kristen Menou\altaffilmark{1},
Caleb A. Scharf\altaffilmark{1,3}}

\affil{$^1$Department of Astronomy, Columbia University, 550 West 120th Street, New York, NY 10027}
\affil{$^2$Department of Astrophysical Sciences, Princeton University, Peyton Hall, Princeton, NJ 08544}
\affil{$^3$Columbia Astrobiology Center, Columbia University, 550 West 120th Street, New York, NY 10027}

\vspace{0.5\baselineskip}

\email{dave@astro.columbia.edu, kristen@astro.columbia.edu,
  caleb@astro.columbia.edu}

\begin{abstract}
Extrasolar terrestrial planets with the potential to host life might
have large obliquities or be subject to strong obliquity
variations. We revisit the habitability of oblique planets with an
energy balance climate model (EBM) allowing for dynamical transitions
to ice-covered snowball states as a result of ice-albedo
feedback. Despite the great simplicity of our EBM, it captures
reasonably well the seasonal cycle of global energetic fluxes at
Earth's surface.  It also performs satisfactorily against a
full-physics climate model of a highly oblique Earth--like planet, in
an unusual regime of circulation dominated by heat transport from the
poles to the equator.  Climates on oblique terrestrial planets can
violate global radiative balance through much of their seasonal cycle,
which limits the usefulness of simple radiative equilibrium
arguments. High obliquity planets have severe climates, with large
amplitude seasonal variations, but they are not necessarily more prone
to global snowball transitions than low obliquity planets. We find
that terrestrial planets with massive $\rm CO_2$ atmospheres,
typically expected in the outer regions of habitable zones, can also
be subject to such dynamical snowball transitions. Some of the
snowball climates investigated for $\rm CO_2$--rich atmospheres
experience partial atmospheric collapse. Since long-term $\rm CO_2$
atmospheric build-up acts as a climatic thermostat for habitable
planets, partial $\rm CO_2$ collapse could limit the habitability of
such planets. A terrestrial planet's habitability may thus depend
sensitively on its short-term climatic stability.
\end{abstract}


\section{Introduction}
\label{obl_sec:intro}

The Earth's obliquity is remarkably stable: the angle between the
spin--axis and the normal to the orbital plane varies by no more than
a few degrees from its present value of $\sim 23.5\degr$.  This
stability is maintained by torque from the Moon
\citep{laskar_et_al1993,nerondesurgy+laskar1997}.  Even within our own
Solar System, though, the obliquity of other terrestrial planets has
varied significantly more; the analysis of \citet{laskar+robutel1993}
indicates that Mars' obliquity exhibits chaotic variations between
$\sim 0\degr$ and $\sim 60\degr$.


How does climate depend on obliquity and its possible variations in
time?  How does the range of orbital radii around a star at which a
planet could support water-based life depend on the planet's
obliquity?  Has the stability of Earth's obliquity made it a more
climatically hospitable home?  The answers to these questions will be
important to evaluate the fraction of stars that have potentially
habitable planets.  There are now more than 300 extrasolar planets
known,\footnote{See http://exoplanet.eu/.} several of which are close
to the terrestrial regime with masses less than 10 times that of the
Earth (e.g., \citealt{beaulieu_et_al2006}, \citealt{udry_et_al2007},
\citealt{bennett_et_al2008}, \citealt{mayor_et_al2008}).  {\it COROT},
which has already launched, and {\it Kepler}, scheduled to launch in
less than one year, are dedicated space-based transit-detecting
observatories that will monitor a large number of stars to detect the
small decreases in stellar flux that occur when terrestrial planets
cross in front of their host stars
\citep{baglin2003,borucki_et_al2003,borucki_et_al2007}.  These
missions are expected to multiply by perhaps several hundredfold or
more the number of known terrestrial planets, depending on the
distribution of such planets around solar-type stars
(\citealt{borucki_et_al2007, borucki_et_al2003, basri_et_al2005};
although see revised predictions in \citealt{beatty+gaudi2008}).  NASA
and ESA have plans for ambitious future missions to obtain spectra of
nearby Earth-like planets in the hope that they would reveal the first
unambiguous signatures of life on a remote world: NASA's {\it
Terrestrial Planet Finder} and ESA's {\it Darwin}
\citep{leger+herbst2007}.  The design of such observatories, and the
urgency with which they will be built and deployed, will depend on the
habitability potential of terrestrial planets that will be found in
the next 5-10 years.

Over the last 50 years, various authors have addressed how to predict
the way in which terrestrial planet habitability depends on
star-planet distance (see \citealt{kasting+catling2003} for a recent
review).  Several of the important initial calculations predated the
first discoveries of extrasolar planets, including \citet{dole1964},
\citet{hart1979}, and the seminal work of \citet{kasting_et_al1993}.
\citet{selsis_et_al2007}, \citet{vonbloh_et_al2008}, and
\citet{barnes_et_al2008} have reconsidered habitability in light of
recent exoplanetary detections.  \citet{williams_et_al1996},
\citet[hereafter WK97]{williams+kasting1997} and
\citet{williams+pollard2003} have tackled precisely the questions
relating to obliquity posed above, and have concluded that variations
in obliquity do not necessarily render a planet non-habitable (see
also \citealt{hunt1982}, \citealt{williams1988b,williams1988c},
\citealt{oglesby+ogg1998}, \citealt{chandler+sohl2000} and
\citealt{jenkins2000,jenkins2001,jenkins2003} in the context of
Earth's paleoclimate studies).

Here we seek to generalize these analyses to model planets that are
less close analogs to Earth than have been considered previously.  In
\citet[hereafter SMS08]{spiegel_et_al2008}, we examined how regionally
and temporally habitable climates are affected by variations in the
efficiency of latitudinal heat transport on a planet, and by
variations in the ocean fraction. Importantly, we found that otherwise
habitable Earth-like terrestrial planets can be subject to dynamical
climate transitions into globally-frozen snowball states. Since it is
not trivial to escape a snowball state (e.g.,
\citealt{pierrehumbert2005}) and such globally-frozen climates may
have profound influences on the development or existence of life
(e.g., \citealt{hoffman+schrag2002}), identifying the likelihood of
such transitions on terrestrial exoplanets should be central to any
habitability assessment.  Following in the footsteps of our first
analysis, here we focus on obliquity and consider the influence on
habitability of several planetary attributes a priori unknown for
exoplanets, such as the efficiency of latitudinal heat transport and
the land-ocean distribution.

The remainder of this paper is structured as follows: In
\S~\ref{obl_sec:model} we describe the energy balance climate model we
use. In \S~\ref{obl_sec:valid} discuss several validation tests in
which our model performs well enough to give us some confidence in its
behavior for conditions that differ from those found on Earth. In
\S~\ref{obl_sec:results} we examine the influence on regional and
seasonal habitability of various excursions from Earth-like
conditions.  Finally, we conclude in \S~\ref{obl_sec:disc+conc}.

\section{Model}
\label{obl_sec:model}

In order to describe the surface temperature and its evolution on a
terrestrial planet, we use a 1-dimensional time-dependent energy
balance model based on a diffusion equation for latitudinal heat
transport.  This type of model has been used in previous
investigations of habitable climates (WK97, SMS08), in modeling
Martian climate under changes in forcing
\citep{nakamura+tajika2002,nakamura+tajika2003}, and in studies of the
Earth's climate (\citealt{north_et_al1981} and references therein).

Our model is based on the following prognostic equation for the
planetary surface temperature (as described in SMS08):
\begin{equation}
\label{obl_eq:diffu eq1}
C \frac{\partial T[x,t]}{\partial t} - \frac{\partial}{\partial x}
\left( D(1 - x^2) \frac{\partial T[x,t]}{\partial x} \right) + I = S
(1 - A).
\end{equation}
In this equation, $x \equiv \sin \lambda$ is the sine of latitude
$\lambda$, $T$ is the temperature, $C$ is the effective heat capacity
of the surface layer, $D$ is the diffusion coefficient that determines
the efficiency of latitudinal redistribution of heat, $I$ is the
infrared emission function (energy sink), $S$ is the diurnally
averaged insolation function (energy source) and $A$ is the
albedo. Our formalism for insolation on oblique planets follows that
of WK97. In the above equation, $C$, $D$, $I$, and $A$ may be
functions of $T$, $x$, $t$, and possibly other relevant parameters.

Our prescriptions for the functions $C$, $D$, $I$, and $A$ follow
SMS08 and are largely borrowed from WK97 or the existing geophysical
literature on similar energy balance models (EBMs). For simplicity and
flexibility, many of our models use very simple, physically motivated
prescriptions. As described in SMS08, we find that an infrared cooling
function of the form
\begin{equation}
I[T] = \sigma T^4 / (1 + (3/4)\tau_{\rm IR}[T])
\label{obl_eq:I}
\end{equation}
(i.e., a one-zone model combined with a simple Eddington transfer
approximation; e.g., \citealt{shu1982}), reproduces the greenhouse
effect on Earth reasonably well for our purposes.  Here, $\tau_{\rm
IR}$ represents the opacity of the atmosphere to long wavelength
infrared radiation.  SMS08 describes three pairs of infrared radiation
functions and albedo functions.  In this analysis, we will use two of
the three: $(I_2,A_2)$, which gives the closest match to the Earth's
annual average temperature distribution, and $(I_3,A_3)$, which uses
the standard linearized cooling function of
\citet{north+coakley1979}. The $(I,A)$ functions are detailed in
Table~\ref{obl_tab:one}.  The albedo functions $A_2$ and $A_3$ are
constant and low ($\sim 0.3$) for high temperatures, constant and high
($\sim 0.7$) for low temperatures (to represent the high albedo of
snow and ice), and vary smoothly in between.  One other $I$ function
that we use is the one proposed in WK97, derived from full
radiative-convective calculations, here denoted $I_{\rm WK97}$.  This
function includes the detailed influence of $\rm CO_2$ (and
implicitely $\rm H_2O$) atmospheric content on radiative cooling.  For
$C$, we assume various configurations of land and ocean, in each
configuration using the same land, ocean, and ice partial $C$ values
as WK97.  Finally, we adopt a diffusion coefficient for latitudinal
heat transport $D_{\rm fid} = 5.394\times 10^2 {\rm
~erg~cm^{-2}~s^{-1}~K^{-1}}\times (\Omega_p/\Omega_\Earth)^{-2}$, as
described in SMS08, where $\Omega_p$ is the angular spin frequency of
the model planet and $\Omega_\Earth$ is that of the Earth. As
explained in SMS08, this scaling largely oversimplifies the complexity
of atmospheric transport expected for planets at different rotation
rates (e.g., \citealt{delgenio_et_al1993,delgenio+zhou1996}).

Equation~(\ref{obl_eq:diffu eq1}) is solved as described in SMS08, on
a grid uniformly spaced in latitude ($1.25\degr$ resolution elements,
found to be sufficient from convergence tests).  We again choose ``hot
start'' ($T\geq 350\rm~K$) initial conditions, to minimize the
likelihood that models will undergo a dynamical transition to fully
ice-covered (snowball) states from which they cannot recover because
of ice-albedo feedback. In that sense, our results on snowball states
are conservative.

To summarize, we make the following assumptions in the models
presented below:
\begin{enumerate}
 \item {\it Heating/Cooling} -- The heating and cooling functions are
    given by the diurnally averaged insolation from a sun--like
    (1~$M_\sun$, 1~$L_\sun$) star, with albedo and insolation
    functions described above and in Table~\ref{obl_tab:one}.
 \item {\it Latitudinal Heat Transport} -- We test the influence on
    climate of three different efficiencies of latitudinal heat
    transport, within the diffusion equation approximation: an
    Earth-like diffusion coefficient, and diffusion coefficients
    scaled down and up by a factor of 9 (which correspond to 8-hour
    and 72-hour rotation according to the above $D\propto
    {\Omega_p}^{-2}$ scaling).
 \item {\it Ocean Coverage} -- We vary both the fraction and the
    distribution of ocean coverage.  For ocean fraction, we present a
    series of models with Earth--like (30\%:70\%) land:ocean fraction, and
    another series of models that represent a desert-world, with a
    90\%:10\% land:ocean fraction.  For ocean distribution, we present
    models in which there is a uniform distribution (in every latitude
    band) of land and ocean, and others in which the land-mass is a
    single continent centered on the North Pole, while the rest of the
    planet is covered with ocean.
 \item {\it Initial Conditions} -- As described in SMS08, the models
    all have a hot-start initial condition, with a uniform surface
    temperature of at least 350~K, to minimize the chances of ending
    up in a globally-frozen snowball state owing solely to the choice
    of initial conditions.  Time begins at the Northern winter
    solstice.
\end{enumerate}

\section{Model Validation}
\label{obl_sec:valid}

In SMS08, we verified that our ``fiducial'' model (70\% ocean;
$I_2,A_2$ cooling-albedo functions) at 1~AU predicts temperatures that
match the Earth's actual temperature distribution at all latitudes
that are not significantly affected by Antarctica (i.e., north of
$60\degr$~S or so).  This indicates that the model accounts for the
overall (annual) planetary energy balance reasonably well.  Another
obvious test is whether the model correctly predicts the monthly
energy fluxes that together go into the overall balance.  Because our
current investigation tests the influence of obliquity on climate, and
obliquity is the primary driver of the Earth's seasons, verifying the
seasonal predictions of our model, given Earth-like conditions, is
particularly relevant.

The diffusion equation model is a statement of conservation of energy.
By definition, after vertical integration for a thin atmosphere with
dominant surface processes,
\begin{equation}
C \frac{\partial T}{\partial t} \equiv \frac{\partial \sigma}{\partial t},
\label{eq:def hc}
\end{equation}
where $\sigma$ is the energy surface density (internal energy per unit
surface area on the globe).  The diffusion equation, therefore, says
that the rate of change of internal energy at a given point equals the
sources of energy (insolation), minus the sinks (infrared radiation),
minus whatever energy flows away from the point under consideration.

\begin{figure}[p]
\plotone
{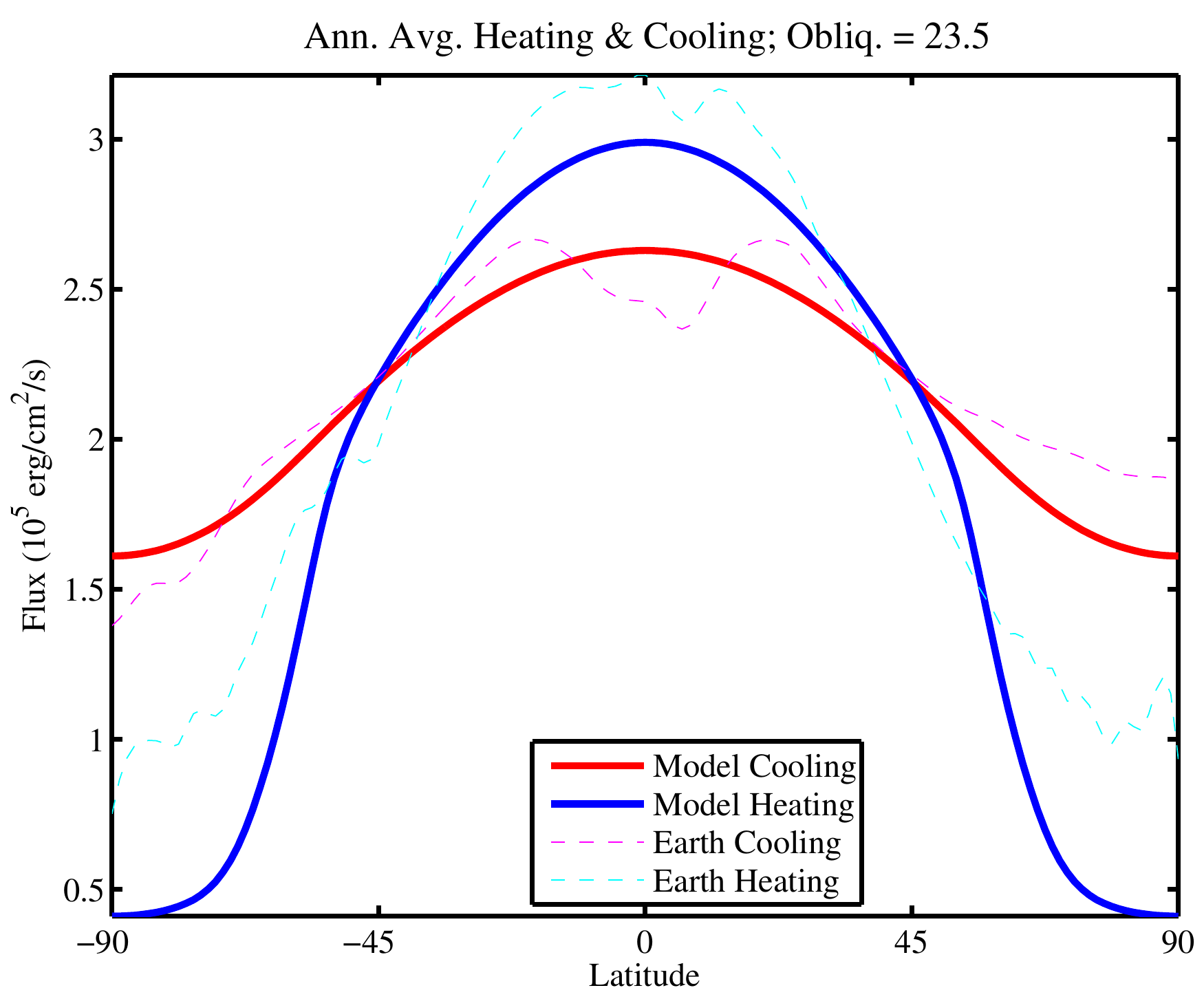}
\caption[Annually averaged cooling and heating fluxes for fiducial
model at 1~AU and Earth.]{Annually averaged cooling and heating fluxes
for our fiducial model at 1~AU and for the Earth. The thick red line
is infrared cooling and the thick blue line is absorbed solar flux, in
our fiducial model (70\% uniform ocean, $I_2,A_2$). The thin dashed
magenta line is the Earth's annually averaged long wavelength infrared
radiation, and the thin dashed cyan line is the annually averaged
absorbed solar flux measured on Earth. Earth-specific features are not
captured by our symmetric and uniform model.}
\label{obl_fig:set30heat_cool_23p5}
\end{figure}

Figure~\ref{obl_fig:set30heat_cool_23p5} presents a comparison between
the annually averaged fluxes of incoming and outgoing radiative energy
in the fiducial model with the corresponding fluxes on Earth, taken
from NASA's Earth Radiation Budget Experiment (ERBE) in the mid-1980s
\citep{barkstrom_et_al1990}.\footnote{The ERBE satellite measured
short-wavelength, or incoming, flux as that from 0.2~$\mu$m to
4.5~$\mu$m.  Long-wavelength, or outgoing, flux was defined as all
other flux within the bolometric range of the instrument.}  While our
model does not capture the full shape of the Earth's cooling and
heating functions -- in particular, the annually averaged model
heating function is a bit below the Earth's at the poles -- still,
both cooling and heating fluxes are within 10\% of the Earth's over
most of the planet's surface.

\begin{figure}[p]
\plotone
{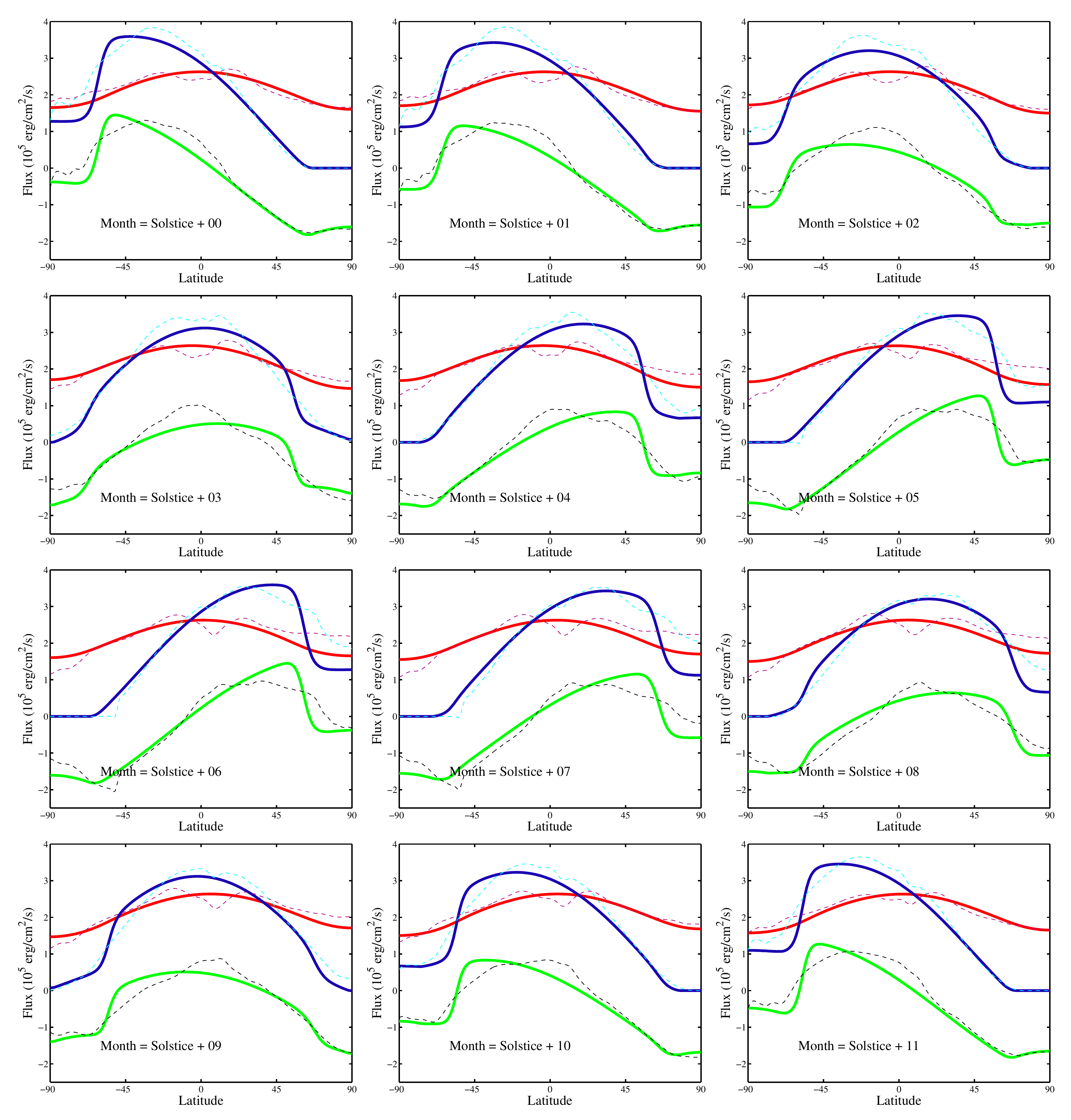}
\caption[Monthly cooling, heating, and net fluxes for fiducial model
at $23.5\degr$ obliquity, at 1~AU, and for Earth.]{Monthly cooling,
heating, and net fluxes for our fiducial model at $23.5\degr$
obliquity and 1~AU orbital distance, and for the Earth.  Each panel
presents the average cooling, heating, and net (heating minus cooling)
radiative fluxes, as functions of latitude, for one month of the year,
starting at the Northern winter solstice (upper left panel), and
incrementing by one month with each panel to the right.  These fluxes
are presented for both the model (thick solid lines) and the Earth
(thin dashed lines).  Model heating is blue; model cooling is red;
model net heating is green.  Earth heating is cyan; Earth cooling is
deep magenta; Earth net heating is black. Our model captures
reasonably well the seasonal variations of these fluxes.}
\label{obl_fig:set30heat_cool_monthly23}
\end{figure}

Figure~\ref{obl_fig:set30heat_cool_monthly23} offers an even more
compelling validation.  In this figure, each of the 12 panels shows
solar (i.e., heating), terrestrial (i.e., cooling), and net (solar
minus terrestrial) radiative fluxes as functions of latitude, averaged
over one month.  Not only are our annually averaged cooling and
heating functions in reasonable agreement with Earth's, as per
Figure~\ref{obl_fig:set30heat_cool_23p5}, but furthermore the temporal
variability of radiative fluxes in our model is similar to that of
Earth.

For example, at the Northern winter solstice (upper left panel of
Fig.~\ref{obl_fig:set30heat_cool_monthly23}), the model heating curve
closely traces that of the Earth.  It peaks at a somewhat more
Southern latitude than the Earth's does, but is within 10\% of the
Earth's at all latitudes north of $60\degr$~S.  As the months advance,
the concordance between the model heating curve and the Earth's
heating curve increases, until there is maximum agreement (within 10\%
at all atitudes) at the equinox (``Solstice+03'').  Then, by the next
solstice, the curves agree to within 10\% at all latitudes South of
roughly $60\degr$~N. In a comparison of the cooling curves, the model
shows even greater agreement with the data.  In a majority of months,
these two curves are within 10\% of each other at all latitudes.

Interestingly, the month-by-month variations in model heating and
model cooling lead to a net heating curve (heating minus cooling) that
predicts some detailed features actually seen in the Earth's net
heating budget.  Notice, for instance, the slight upward turn of the
net heating curves of both the model and the Earth near the North
Pole, at and around the Northern winter solstice.  A similar feature
is seen in both curves (though with slightly less impressive detailed
agreement) near the South Pole, at and around the Southern winter
solstice.  These comparisons establish that our climate model exhibits
reasonable regional and seasonal variability of not just temperature
but also incoming and outgoing radiative energy fluxes.

\begin{figure}[p]
\plotone
{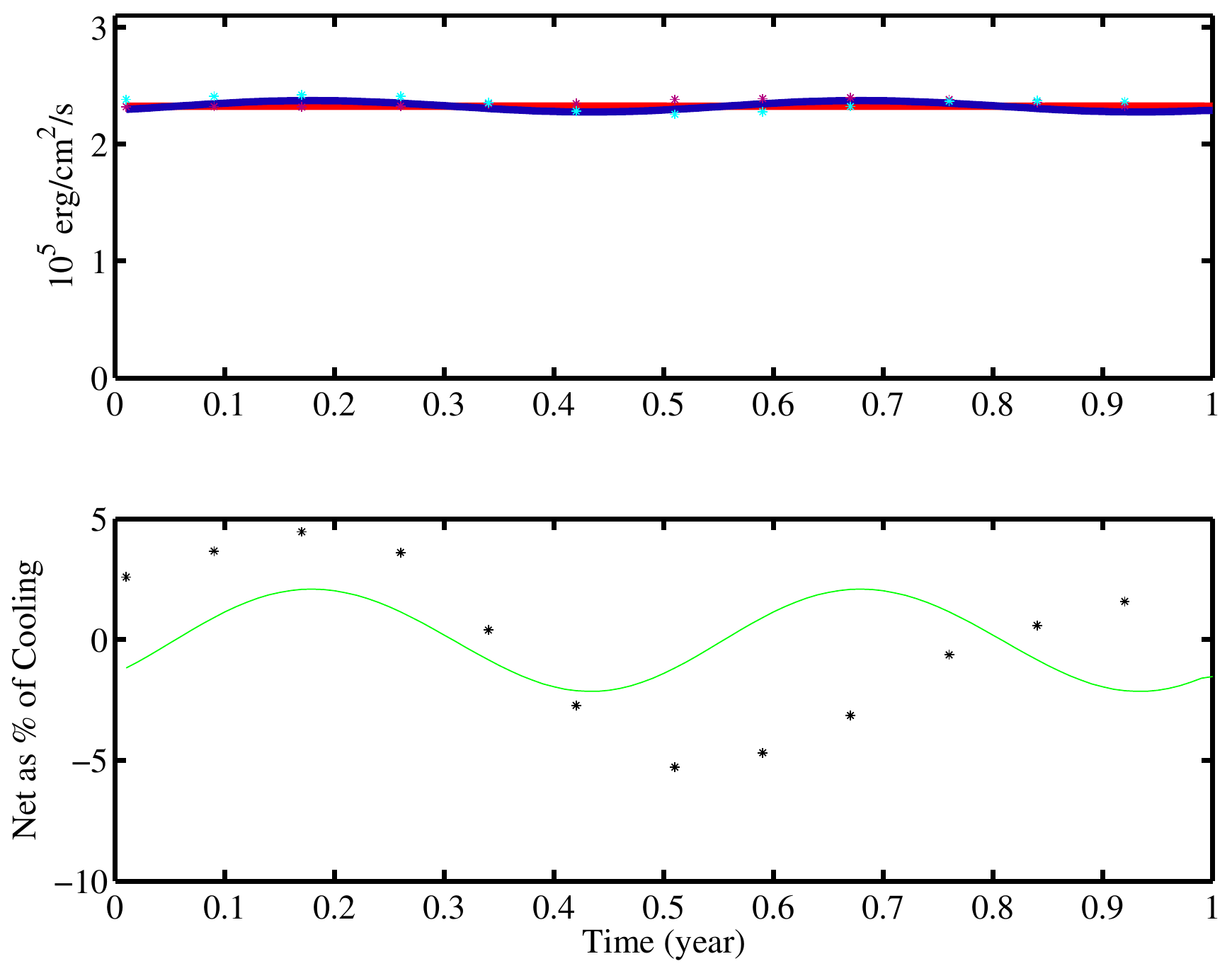}
\caption[Global average cooling, heating, and net radiative flux, as
functions of time in fiducial model at $23.5\degr$ obliquity.]{Global
average cooling, heating, and net radiative flux, as functions of time
in our fiducial model at $23.5\degr$ obliquity (solid lines). Earth's
ERBE data are shown as stars. {\it Top Panel:} Global average cooling
(red curves and magenta stars) and heating (blue curves and cyan
stars) fluxes as a function of time of year, measured in fraction of a
year from the Northern winter solstice.  {\it Bottom Panel:} Net
heating flux (heating minus cooling) for the model (green curve) and
ERBE data (black stars), plotted as percent of the corresponding
cooling flux. The Earth remains within $5 \%$ of global radiative
balance throughout its seasonal cycle.}
\label{obl_fig:set30_heat_cool_vs_time}
\end{figure}

Another way to consider seasonal variations of heating and cooling
fluxes is to look at the globe--average of each with respect to time.
Figure~\ref{obl_fig:set30_heat_cool_vs_time} presents a comparison of
these fluxes, for our fiducial model and the Earth itself.  The bottom
panel of this figure shows the net heating flux as a function of time
of year, measured in fraction of a year from the Northern winter
solstice.  Earth's net heating flux varies by about 5\% with respect
to the cooling flux, while our model's varies somewhat less.  The
heating function for the Earth exceeds the cooling function during
Northern winter for two main reasons:\footnote{Note that the cooling
function, which traces surface temperatures, varies less through the
seasonal cycle than the heating function.} First, the nonzero
eccentricity ($e\approx 0.0167$) of the Earth's orbit places its
perihelion -- which occurs during Northern winter -- approximately
3.4\% closer to the Sun than its aphelion.  This is responsible for
$\sim 7\%$ of the annual $\sim 10\%$ annual variation in net heating
flux. Our fiducial model, on the other hand, assumes zero
eccentricity. Another contributing factor is that the Earth's oceans
on average absorb somewhat more insolation than the land, and the
Southern hemisphere -- which faces the Sun during Northern winter --
has greater ocean coverage than the Northern hemisphere. To within $5
\%$, however, the Earth remains in global radiative balance throughout
its seasonal cycle.

So far we have considered the radiative fluxes, but what about
diffusive energy flux?  We may combine equation~(\ref{obl_eq:diffu
eq1}) with equation~(\ref{eq:def hc}) to produce:
\begin{equation}
\frac{\partial \sigma}{\partial t} - \frac{\partial}{\partial x} \left\{ D\cos^2\lambda \frac{\partial}{\partial x}T \right\} = \left({\rm sources - sinks}\right)_{\rm energy~per~area},
\label{eq:act. eq. sig}
\end{equation}
where we have substituted $\cos^2 \lambda$ for $(1-x^2)$.  Comparing
equation (\ref{eq:act. eq. sig}) to a diffusion equation in spherical
coordinates, and accounting for vertical integration, shows that
$F_\lambda = R D \cos\lambda (\partial T/\partial x)$ is the rate of
latitudinal energy transport per unit longitudinal length.  The total
rate of meridional diffusive heat transport (i.e. energy crossing a
given latitude circle per unit time) therefore is
\begin{equation}
\mathcal{F}_\lambda = 2\pi R \cos\lambda F_\lambda = 2 \pi R^2 D
\cos^2\lambda \frac{\partial T}{\partial x}.
\label{eq:total flux}
\end{equation}

\begin{figure}[p]
\plotone
{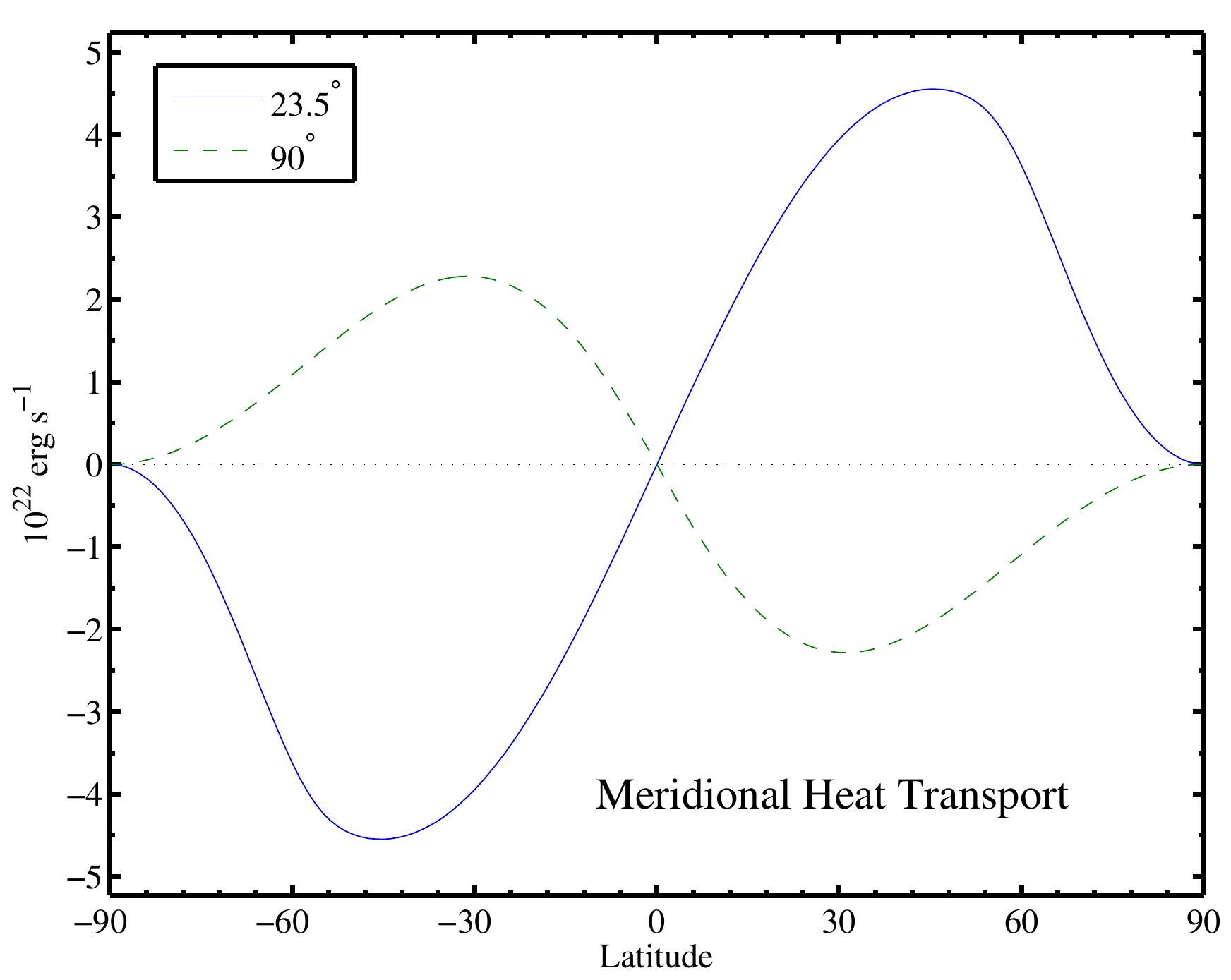}
\caption[Annually averaged, longitudinally integrated, meridional heat
transport at $23.5\degr$ and $90\degr$ obliquity.]{Annually averaged,
longitudinally integrated, meridional heat transport rates in models
at $23.5\degr$ and $90\degr$ obliquity.  Transport is positive
northward.  In the Earth-like case (blue solid curve), heat flows from
the equator to the poles; in the highly oblique case (green dashed
curve), the annually averaged heat flow is reduced and in the opposite
direction. This result is in close agreement with a comparable one
obtained by \citet{williams+pollard2003} with a full physics climate
model.}
\label{obl_fig:set80merid_heat_flux}
\end{figure}

Figure~\ref{obl_fig:set80merid_heat_flux} shows profiles of this
diffusive heat transport rate in our fiducial model, at Earth-like
$23.5\degr$ obliquity and at extreme $90\degr$ obliquity.  In the
Earth-like configuration, heat flows from the equator toward the
poles.  In the highly oblique configuration, however, heat flows in
the other direction, from the poles to the equator (in an annually
averaged sense). For comparison, \citet{williams+pollard2003} present
a full general circulation model (GCM) of an Earth-like planet at
Earth-like and higher obliquity.  Figure~2 of that paper shows the
meridional heat flux within their models for $23.5\degr$ obliquity and
$85\degr$ obliquity, and the results are strikingly similar to ours.
At $23.5\degr$ obliquity, our model's diffusive flux is very close to
that of the GCM.  At high obliquity, the flux in our model remains
within $\sim 30\%$ of that in the GCM (from visual inspection), at all
latitudes.  This reasonable concordance indicates that the treatment
of heat transport within our model, despite being very simple, is
still likely to remain useful as a representation of heat transport in
less-Earth-like conditions.  We emphasize that it is a nontrivial
point that this entirely different regime of transport should remain
well-captured by a diffusion approximation.

\section{Study of Habitability}
\label{obl_sec:results}

For model planets with $23.5\degr$ obliquity on a circular orbit at
1~AU, both pairs of infrared cooling functions and albedo functions
presented in Table~\ref{obl_tab:one} are reasonably good matches for
the Earth's current climate, as measured by latitudinally averaged
temperatures, with a somehwat better fit with $(I_2,A_2)$.  This gives
us some confidence that these functions are useful guides as to how
the climate might respond under different forcing conditions.  In this
investigation, we consider how variations in intrinsic planetary
characteristics combine with the changes in insolation and year-length
at various orbital radii to map the zone of regionally habitable
climates on planets with various obliquities.

We follow SMS08 in saying that, at a given time, a part of a planet is
habitable if its surface temperature is between 273~K and 373~K,
corresponding to the freezing and boiling points of pure water at
1~Atm pressure.  This criterion may be criticized for several reasons
discussed in SMS08 and references therein, but it provides a
reasonable starting point for making numerical investigations.  We
will frequently quantify habitability of pseudo-Earths with the
temporal habitability fraction, $f_{\rm time}[a,\lambda]$, where $a$
is orbital semimajor axis, $\lambda$ is latitude, and $f_{\rm time}$
is the fraction of the year that the point in parameter space
specified by $(a,\lambda)$ spends in the habitable temperature range
(see SMS08 for details).

\subsection{Efficiency of Heat Transport}
\label{obl_ssec:efficiency}

Terrestrial planets with different rotation rates will redistribute
heat from the substellar point (or, in a 1D model, the substellar
latitude) with different efficiencies.  According to the idealized
scaling described above, wherein the effective diffusion coefficient
varies with the inverse square of the planetary rotation rate, slower
spinning planets will redistribute heat more efficiently, while faster
spinning planets will do so less efficiently.  But from where, and to
where, is heat redistributed?  How does this depend on obliquity and
rotation rate?  And what influence does this have on climatic
habitability?

\subsubsection{Direction of Heat Flow}
\label{obl_sssec:direction}

For an Earth-like $23.5\degr$ obliquity, the substellar latitude does
not vary very much over the course of the year: the tropics are fairly
close to the equator (the tropical region is less than one third of
the Earth's surface area).  As a result, it is a reasonable
approximation that heat is always being transported from the equator
to the poles (but see Fig.~\ref{obl_fig:set30heat_cool_monthly23} for
details).  In contrast, on a planet with significantly larger
obliquity, the direction of heat flow changes over the course of the
annual cycle.  At the equinoxes, the equator is the most strongly
insolated part of the planet (regardless of obliquity), and so heat
builds up at the equator, to be partially redistributed by atmospheric
motions.  But on a highly oblique planet, polar summers are extremely
intense, as measured by diurnally averaged insolation.  As a result,
heat builds up at the poles during their corresponding summers, and
the flow of heat reverses direction.

\begin{figure}[p]
\plottwo
{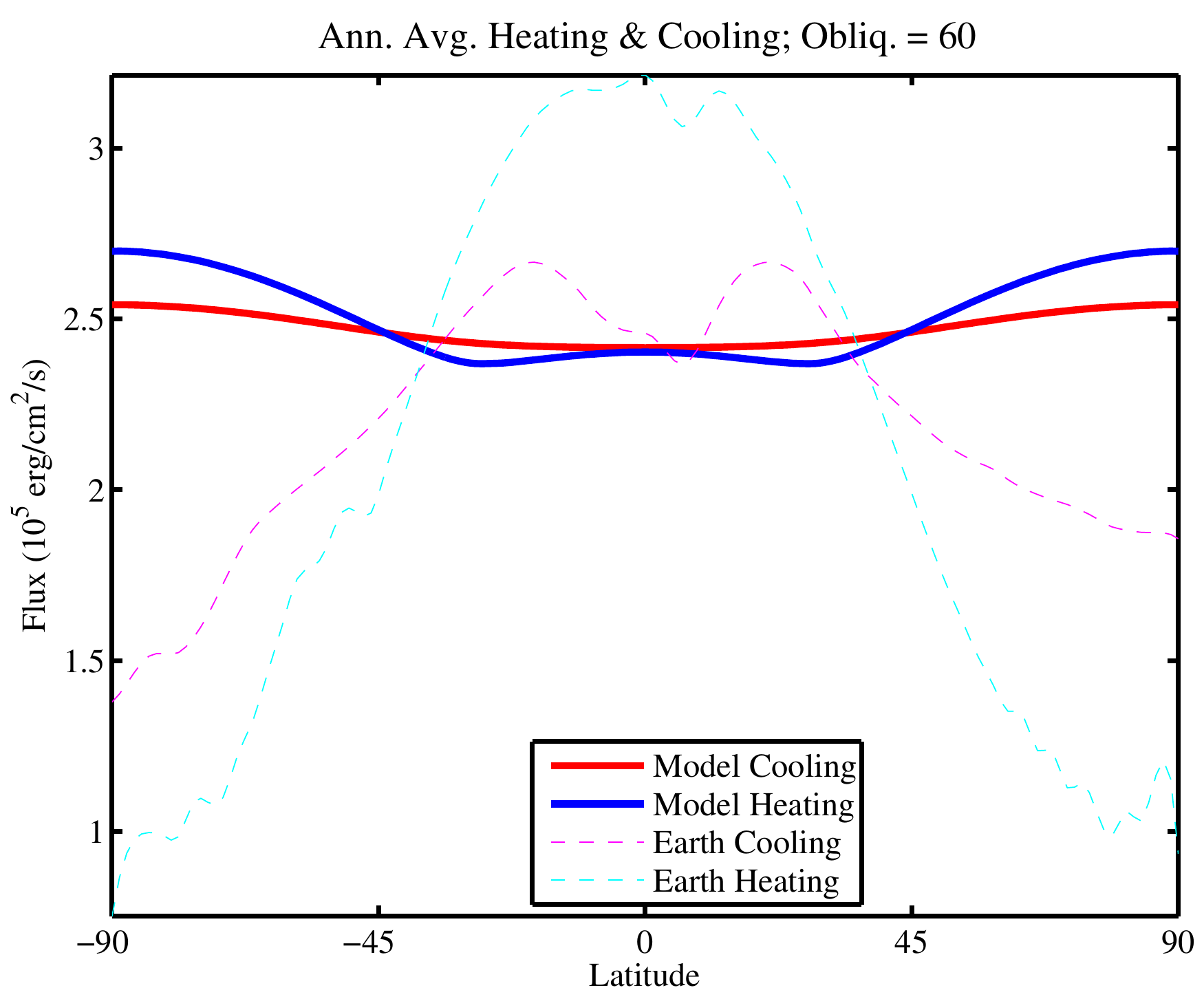}
{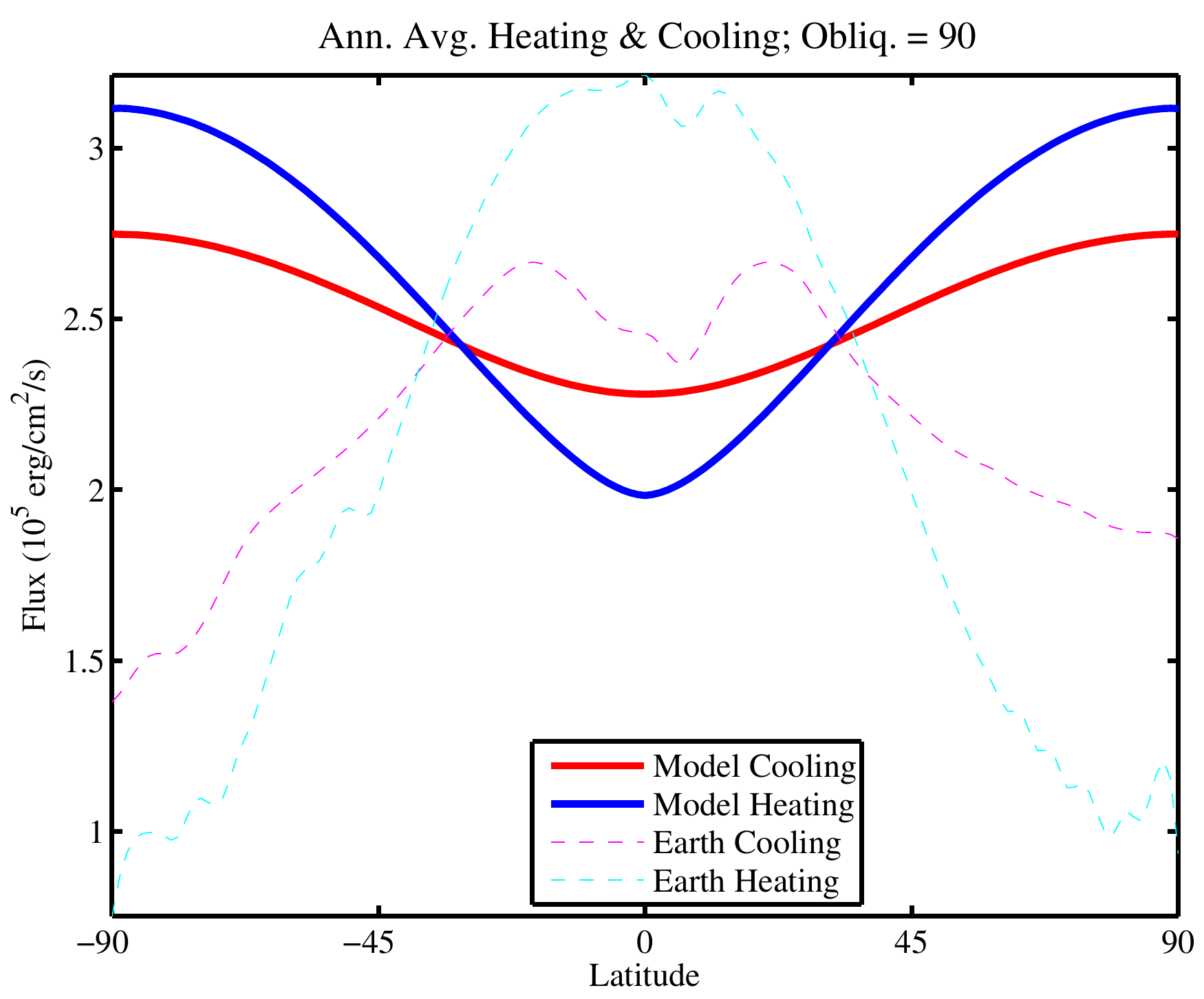}
\caption[Annually averaged cooling and heating fluxes for high and
extreme obliquity model planets model and for Earth.]{Annually
averaged cooling and heating fluxes for high and extreme obliquity
model planets, compared to Earth.  By contrast with
Fig.~\ref{obl_fig:set30heat_cool_23p5}, which shows the annually
averaged heating and cooling for $23.5\degr$ obliquity, here we
present these functions for Earth-like models at $60\degr$ obliquity
(left panel) and at $90\degr$ obliquity (right panel). Thick lines are
model results; thin dashed lines are Earth's ERBE data.  In both high
obliquity cases, unlike on Earth, there is net annually averaged
heating at the poles (i.e., heating exceeds cooling), and, especially
for the extreme obliquity case, net annually averaged cooling at the
equator.}
\label{obl_fig:set30heat_cool_6090}
\end{figure}

Figure~\ref{obl_fig:set30heat_cool_6090} demonstrates the effect of
such strong polar summers on the global radiation budget of a model
planet, by comparison to Figure~\ref{obl_fig:set30heat_cool_23p5} in
which the annually averaged cooling and heating are shown for an
Earth-like, $23.5\degr$ obliquity model.  As expected for the
Earth-like model, over the annual cycle, the equator receives
significantly more solar radiation than do the poles, and accordingly
the annually averaged heating exceeds the cooling at the equator.
This indicates that atmospheric motions transport heat poleward from
the equator on average.  In Figure~\ref{obl_fig:set30heat_cool_6090},
on the other hand, we present the analogous functions in the case of
high and extreme obliquity models. The left panel shows the heating
and cooling functions for a model at $60\degr$ obliquity; the right
panel shows the same functions for a model at $90\degr$ obliquity.
Planetary scientists have long recognized that in highly oblique
models such as these, the polar summers are so intense that, averaged
over the year, the most strongly insolated parts of the planet are the
North and South Poles!  (See, e.g., \citealt{ward1974}.)  In an
annually averaged sense, then, heat flows from the poles to the
equator, although clearly the direction of flow changes with the
seasons, as described above.  The import of these plots is that our
notion that the poles are the coldest planetary regions might have to
be revised in the case of highly oblique worlds. The resulting regime
of atmospheric transport, which is only parameterized with our
diffusive treatment, may also be expected to differ substantially from
that on Earth (e.g., in terms of equatorial Hadley cells).

\begin{figure}[p]
\plotone
{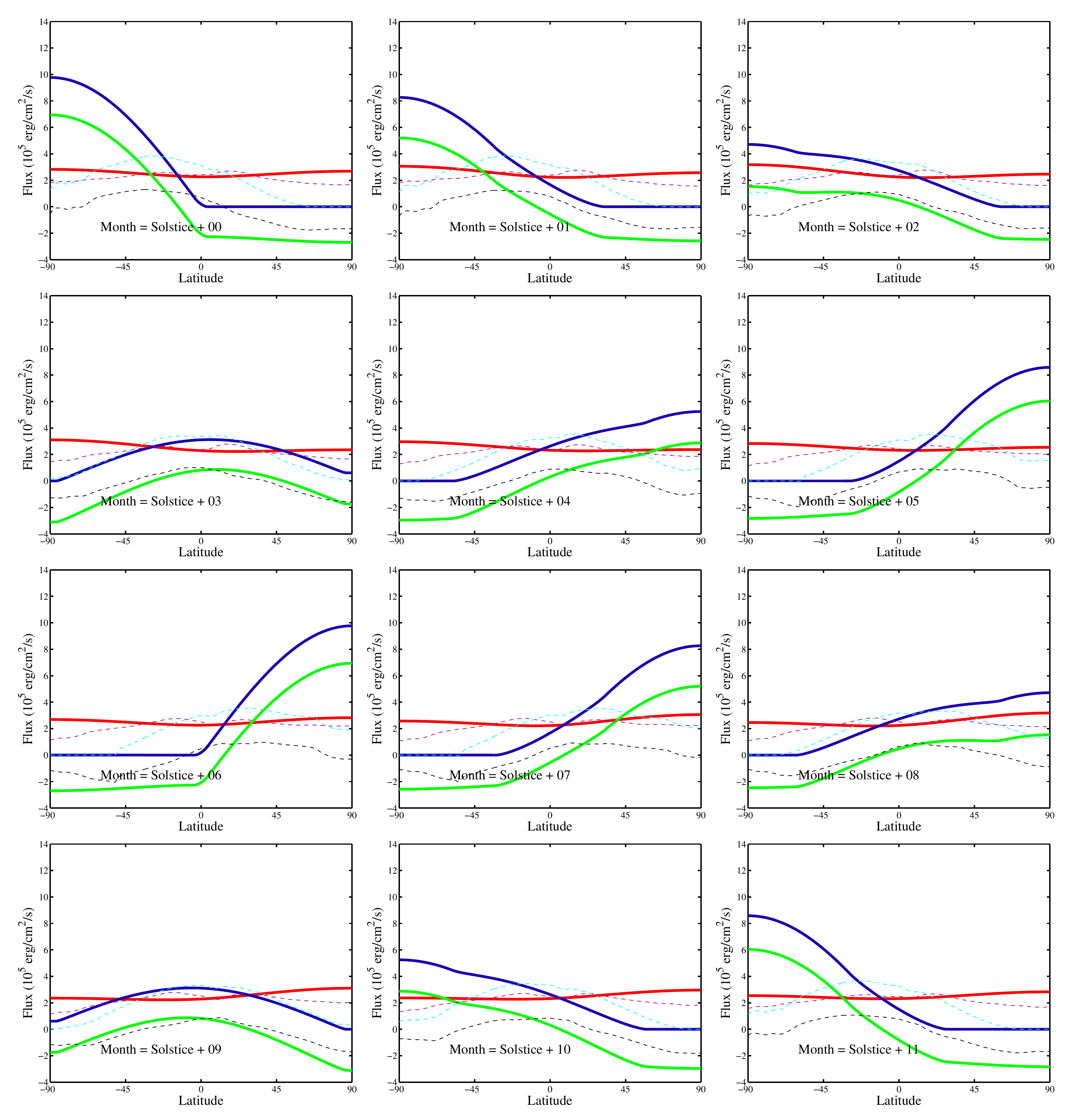}
\caption[Monthly cooling, heating, and net fluxes for fiducial model
at $90\degr$ obliquity, at 1~AU, and for Earth.]{Monthly cooling,
heating, and net heating fluxes for the fiducial model at $90\degr$
obliquity and 1~AU orbital distance, and for the Earth.  Each panel
presents the average cooling, heating, and net (heating minus cooling)
radiative fluxes, as functions of latitude, for one month of the year,
starting at the Northern winter solstice (upper left panel), and
incrementing by one month with each panel to the right.  These fluxes
are shown for both the oblique model planet (thick solid lines) and
the Earth (thin dashed lines).  Model heating is blue; model cooling
is red; model net heating is green.  Earth heating is cyan; Earth
cooling is deep magenta; Earth net heating is black. While cooling
fluxes remain relatively steady and uniform on such an oblique planet,
heating and net heating fluxes are subject to very large seasonal
variations.}
\label{obl_fig:set30heat_cool_monthly90}
\end{figure}

Figure~\ref{obl_fig:set30heat_cool_monthly90} shows in greater detail
the extreme way in which insolation can vary over the annual cycle in
a highly oblique model.  In this model, the obliquity of an otherwise
Earth-like planet ($I_2$--$A_2$ functions, with 70\% ocean uniformly
distributed) is set to 90$\degr$.  Notice that the cooling remains
much more steady than the heating in this model.  This is because of
the high effective heat capacity of the atmosphere above ocean.  In
models with less ocean coverage, or oceans that are nonuniformly
distributed, the cooling too can vary dramatically over the annual
cycle.\footnote{It is also worth noting that at the outer reaches of a
$\sim 2M_\sun$ star's habitable zone
\citep[e.g.,][]{kasting_et_al1993}, the annual cycle might be long
enough relative to the thermal timescale ($\sim$ a decade) of the
ocean-atmosphere mixed layer that it could undergo relatively large
swings in temperature within a single annum.}  Because the heating
varies so intensely while the cooling varies less, their difference --
the net heating curve -- also exhibits large variations within the
annual cycle.  At each solstice, the pole facing the star receives far
more net radiant flux than both the opposite pole and the planetary
mean.  At the equinoxes, something perhaps more surprising happens:
the net radiant flux is negative over much of the model planet's
surface, and only barely positive near the equator.  Overall, the
planet heats strongly at the poles during solstices (while cooling
elsewhere) and either cools or remains essentially thermally neutral
everywhere during equinoxes.


\subsubsection{Implications for Habitability}
\label{obl_sssec:hab}

As SMS08 demonstrates, model planets with efficient heat transport
(slowly spinning planets) are more latitudinally isothermal than
models with Earth-like rotation, which themselves exhibit less
latitudinal variation of temperature than those with inefficient heat
transport (fast spinning planets).  As a result, models
corresponding to worlds that are spinning slowly relative to the Earth
(but still fast enough that a 1-D climate model has some use) tend to
be either entirely habitable or entirely non-habitable at any given
time.  In contrast, Earth-like and faster spinning model planets may
be only partially habitable at a particular time.  They may, for
instance, be frozen at the poles and temperate at the equator, or vice
versa in the case of a highly oblique world.

\begin{figure}[p]
\plotone
{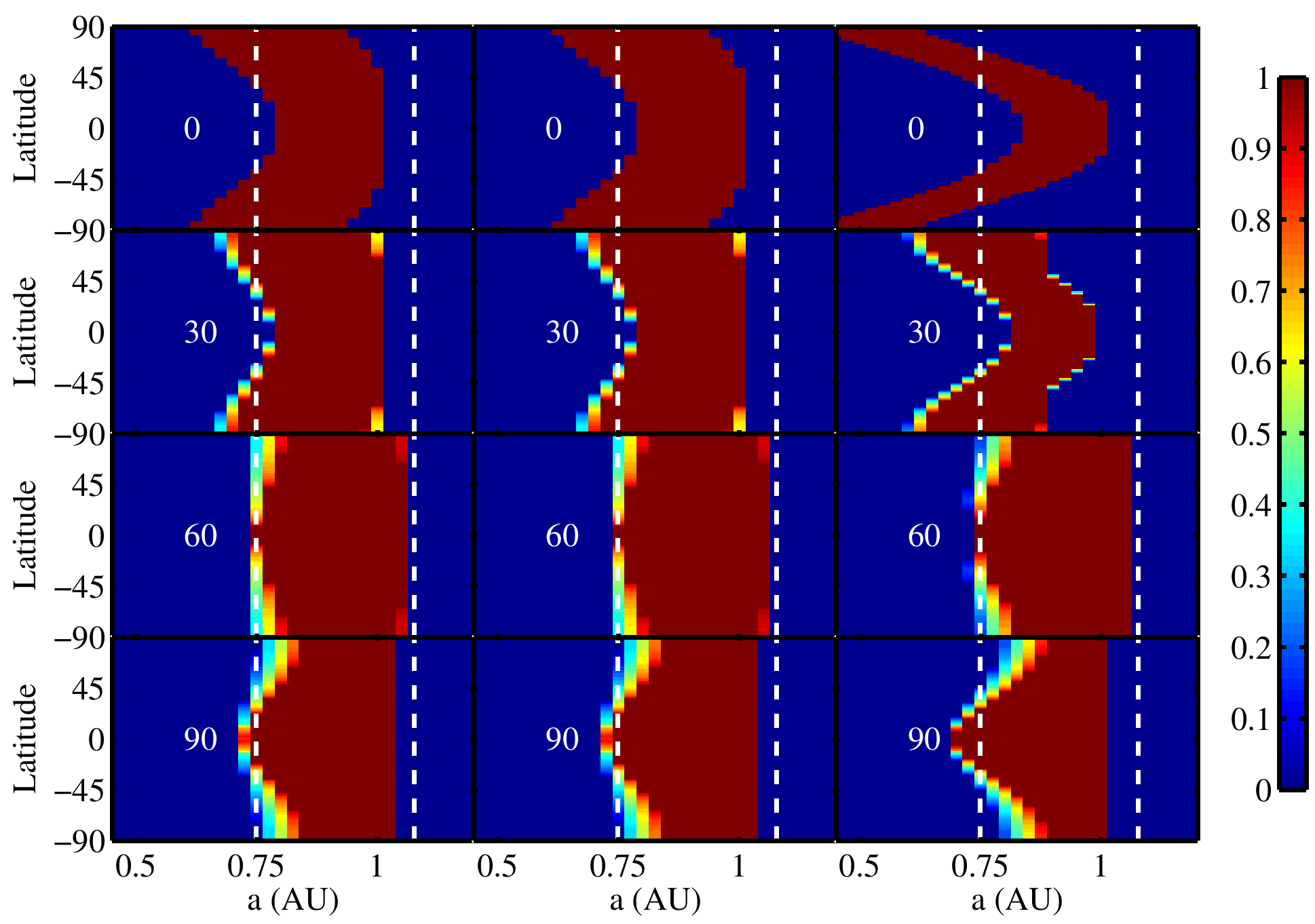}\\
\plotone
{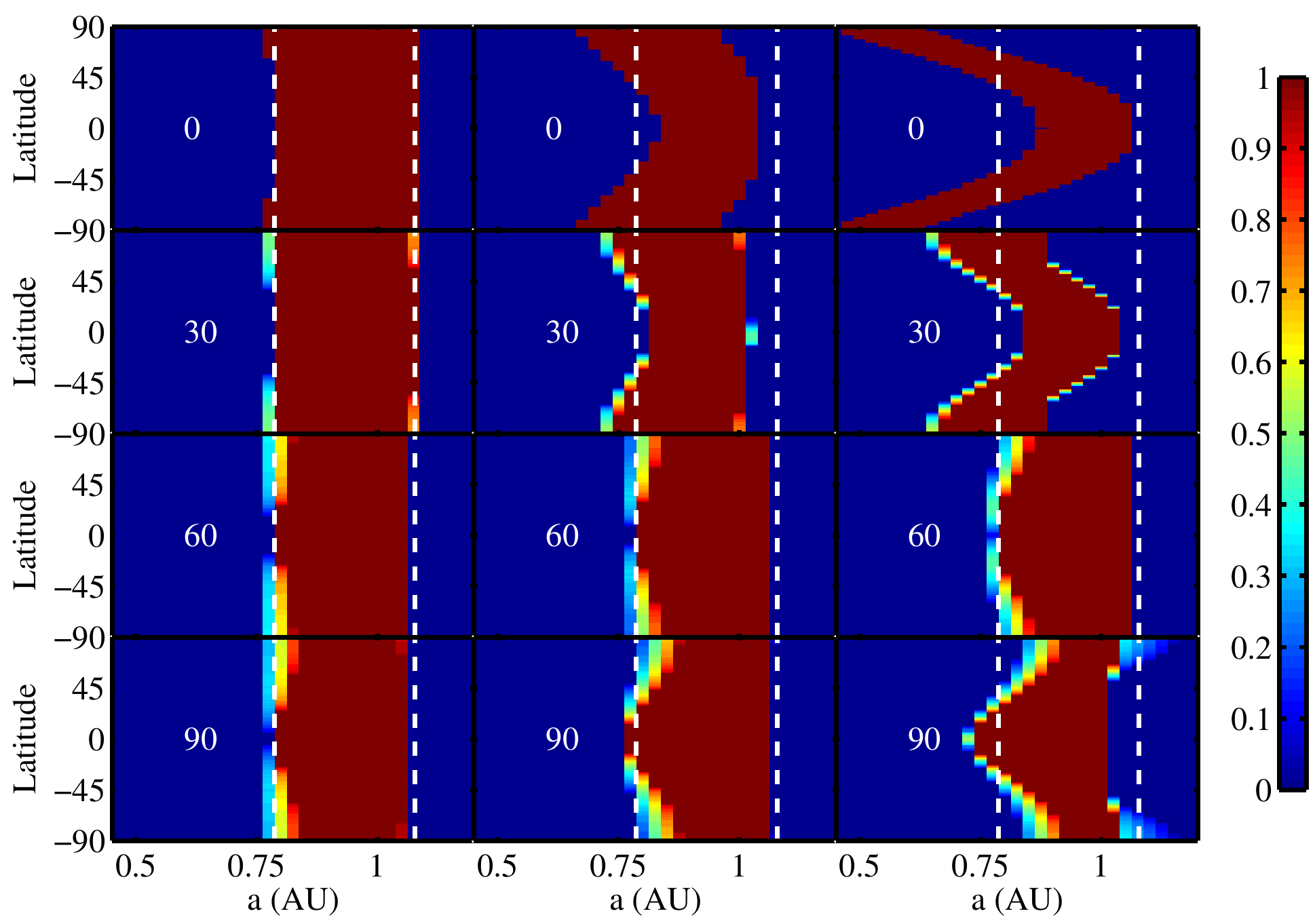}
\caption[Model temporal habitability fraction under different
obliquities, rotation rates, and cooling/heating
functions.]{\footnotesize Temporal habitability fraction in models
with different obliquities, heat transport efficiencies, and
cooling-albedo functions. In both panels, obliquity varies from
$0\degr$ (top row) to $90\degr$ (bottom row). The latitudinal heat
diffusion coefficient varies from $D_{\rm fid}/9$ (left column) to
$D_{\rm fid}$ (center) and $9D_{\rm fid}$ (right).  In each panel, the
abscissa is orbital radius, in AU, and the ordinate is latitude.
Colors indicate the fraction of the year spent by that region in the
habitable temperature range (273~K - 373~K).  {\it Top Panel:}
($I_2,A_2$) cooling-albedo combination. {\it Bottom Panel:}
($I_3,A_3$) cooling-albedo combination. For comparison, vertical
dashed lines show the habitable zone extent expected on the basis of
global radiative balance. Regionality and seasonality can extend the
inner reach of the instantaneous habitable zone while often reducing
its outer reach, when global snowball transitions occur.}
\label{obl_fig:obliq_rotI2A2I3A3}
\end{figure}

Figure~\ref{obl_fig:obliq_rotI2A2I3A3} demonstrates the complicated
interplay that can go on between obliquity and efficiency of heat
transport, in determining a planet's habitability.  This figure shows
the temporal habitability fraction, as a function of orbital semimajor
axis and latitude, for each of 12 different combinations of obliquity
($0\degr$, $30\degr$, $60\degr$, $90\degr$) and latitudinal heat
diffusion coefficient ($D_{\rm fid}/9$, $D_{\rm fid}$, and $9D_{\rm
fid}$, corresponding respectively to 72-hour, 24-hour, 8-hour
rotations).  The top panel shows these plots for the ($I_2,A_2$)
cooling--albedo pair, and the bottom panel depicts the ($I_3,A_3$)
pair (see Table~\ref{obl_tab:one} for details).  In both panels, the
left column of plots represents efficient latitudinal heat transport;
the middle column represents Earth-like transport; and the right
column represents inefficient transport.  Each of the 24 plots in this
figure shows the results of model runs for planets located from
$0.45$~AU to 1.25~AU in increments of $0.025$~AU. The color scale
indicates the fraction of the year that the latitude at that point
spends in the habitable temperature range (273~K - 373~K) on a model
planet at the specified orbital semimajor axis.  In each plot, the
white vertical dashed lines indicate the radiative equilibrium
habitable zone, calculated (as discussed in SMS08) from a
0-dimensional model with annually averaged, globally averaged
insolation and cooling.

There are a number of interesting features in
Figure~\ref{obl_fig:obliq_rotI2A2I3A3}.  The most obvious one is that,
as expected, at every obliquity, less efficient transport results in
more strongly latitudinally differentiated temporal habitability.  In
addition, at each transport-efficiency value, the {\tt $>$}~sign shape
of the seasonally habitable ribbon at low obliquity reverses to a {\tt
$<$}~sign shape at high obliquity.  In other words, at low obliquity,
the relatively cold poles are habitable closer to the star and the
relatively warm equator is habitable farther from the star. At high
obliquity, however, this reverses and the poles are relatively warm,
while the equator is comparatively colder.

Furthermore, in both panels, the contours in most cells show a very
abrupt outer boundary to the seasonally habitable zone.  This is
because, as discussed in SMS08, the ice-albedo feedback renders these
models quite sensitive to changes in forcing.  Small reductions in
insolation can be amplified, because the ice-coverage increases, which
increases the global albedo and leads to further reduction in
insolation.  This feedback mechanism renders the model climates
susceptible to a global snowball transition, from which they cannot
recover within our model framework.  \footnote{Additional feedback
mechanisms that are not incorporated in our model exist on a real
planet and might help it to recover from a snowball state.  For
discussions of how the Earth might have recovered from one or more
snowball episodes, see, for example, \citet{caldeira+kasting1992},
\citet{hoffman+schrag2002}, and \citet{pierrehumbert2004}.}  The main
exceptions to this trend are the low obliquity, fast-spinning models,
in the upper right corners of both panels, although even these models
drop to 0\% habitability at orbital radii that are small relative to
the outer boundary of the habitable zone set by global radiative
equilibrium (indicated by the white dashed lines).  Interestingly, the
fairly small difference in cooling--albedo functions from ($I_2,A_2$)
to ($I_3,A_3$) is sufficient to allow the cell in the lower right
corner of the bottom panel -- extreme obliquity, inefficient transport
-- to avoid transitioning globally to a snowball state.  In that
model, the intense summer insolation at the poles, combined with the
relative thermal isolation of different latitudes, allows the poles to
heat up above the freezing point of water during their summers, even
at orbital distances where other models would be entirely frozen.  In
sum, susceptibility to snowball transitions depends on details of
parameterizations in our energy balance model, as has been noted
before for Earth climate studies \citep[e.g.,][]{north_et_al1981}.

\subsection{Land/Ocean Distribution}
\label{obl_ssec:ocean distribution}

As described in SMS08, the large covering fraction of oceans on the
Earth (roughly 70\%) stabilizes our climate over an annual cycle, by
virtue of the large effective heat capacity of atmosphere over ocean.
Over land, the thermal relaxation timescale is several months, while
for the ($\sim 50$~m deep) mixed layer over the ocean, the thermal
relaxation timescale is more than a decade.  As a result, in a 1D
model (such as ours) that does not resolve continents in longitude,
any latitude band with significant ocean fraction will have strongly
suppressed annual temperature variations relative to a latitude band
with low ocean fraction.  Because we do not know of any way to
determine a priori the distribution of continents and oceans on an
extrasolar planet, it is important to consider the influence on
climatic habitability of other possible land/ocean distributions.

\subsubsection{Nonuniform Ocean Coverage}
\label{obl_sssec:nonuniform}

We consider model planets with distributions of land and ocean that
are not uniform across different latitudes: one with 30\% land
coverage, with a land-mass centered on the North Pole (extending down
to $\sim 24\degr$ North latitude), and the other (discussed in
\S~\ref{obl_sssec:desert}) with 90\% land, again centered on the North
Pole.\footnote{Equivalently, this model can be conceived as having an
ocean centered at the South Pole.}  Because of the relatively low
thermal inertia of atmosphere over land, parts of a model planet that
are dominated by land can freeze or boil during the course of the year
and still return to temperate conditions at other times.  In fact, at
some orbital distances, and at high obliquity, the polar regions of
some models freeze {\it and} boil within an annual cycle.

\begin{figure}[p]
\plottwo
{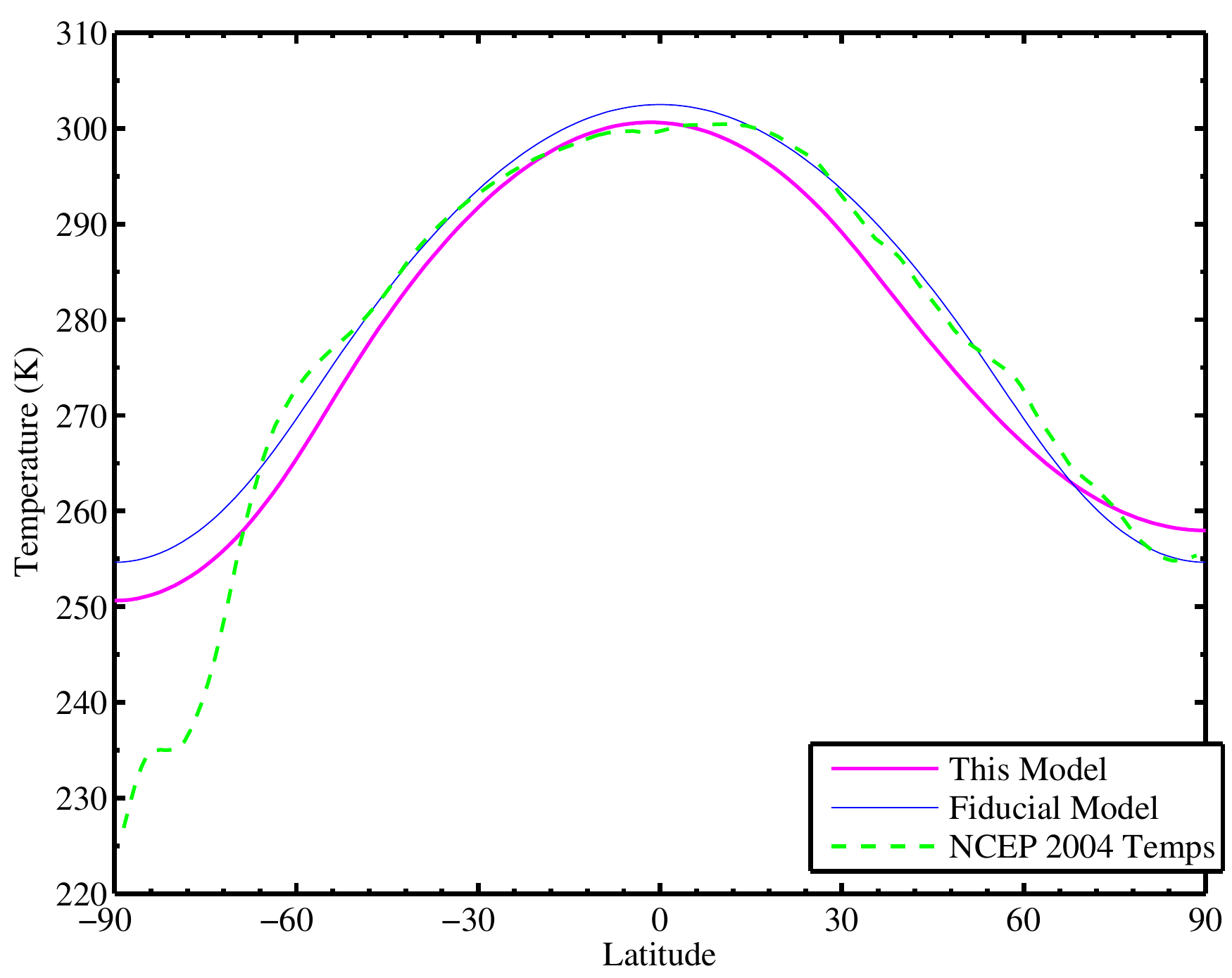}
{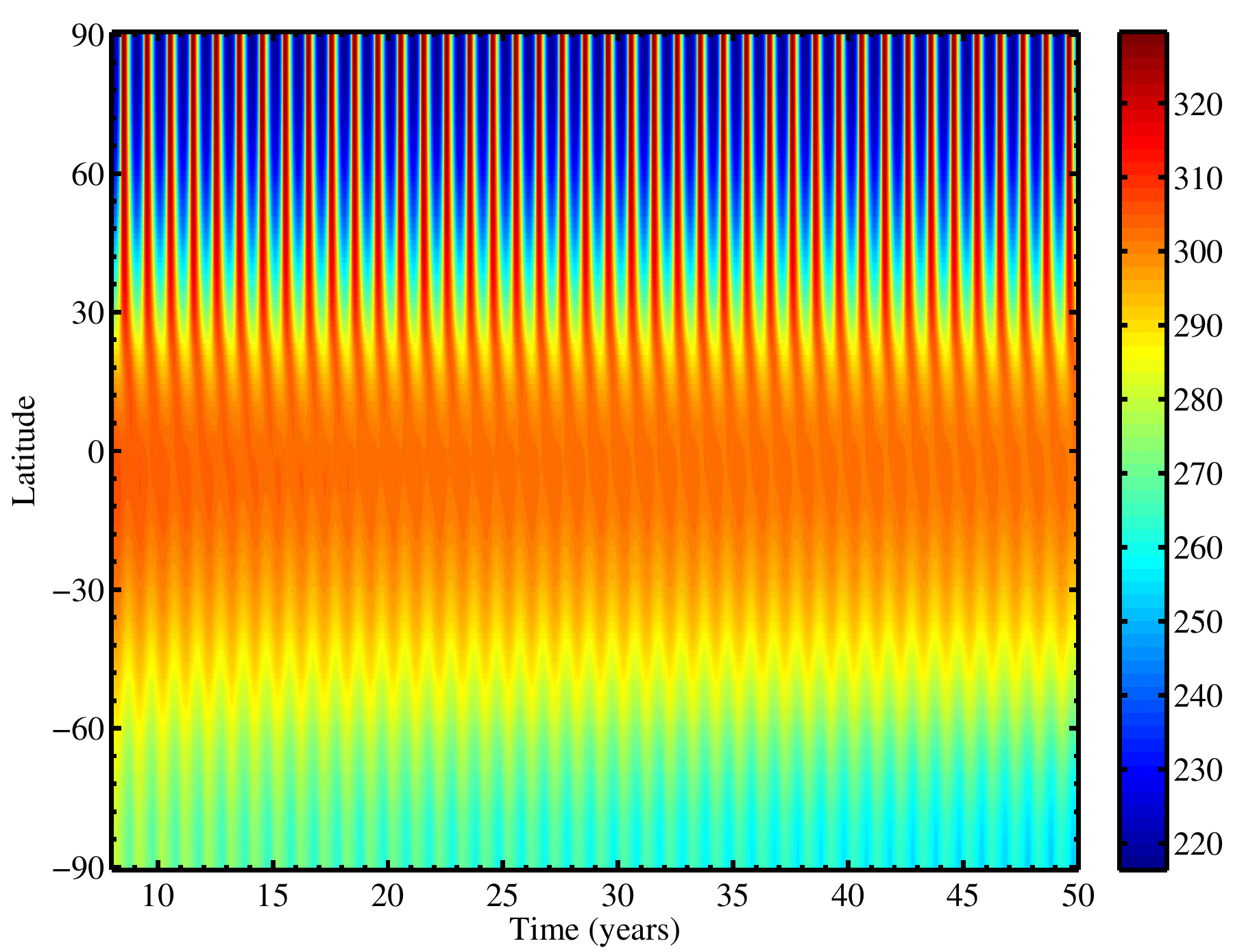}\\
\plottwo
{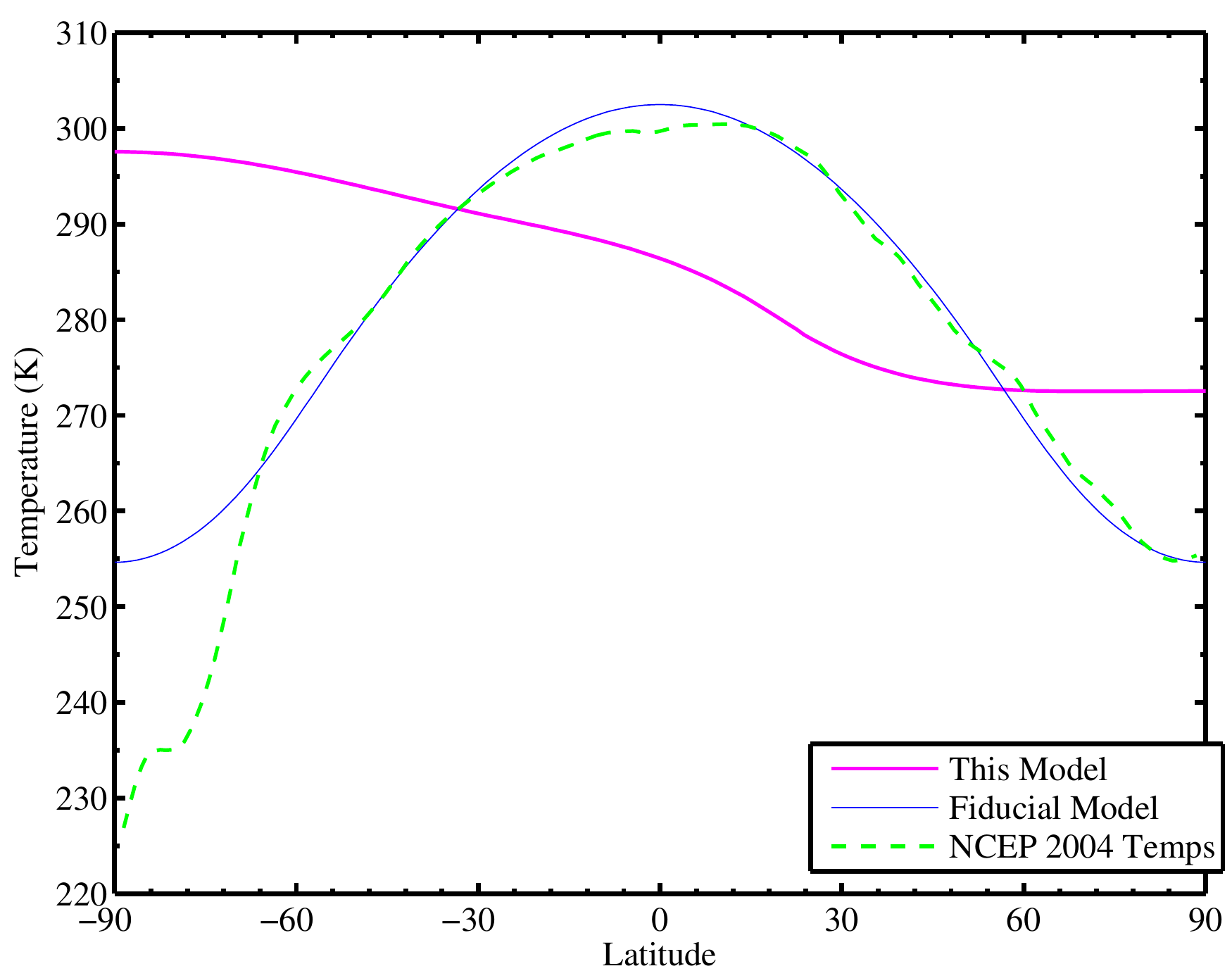}
{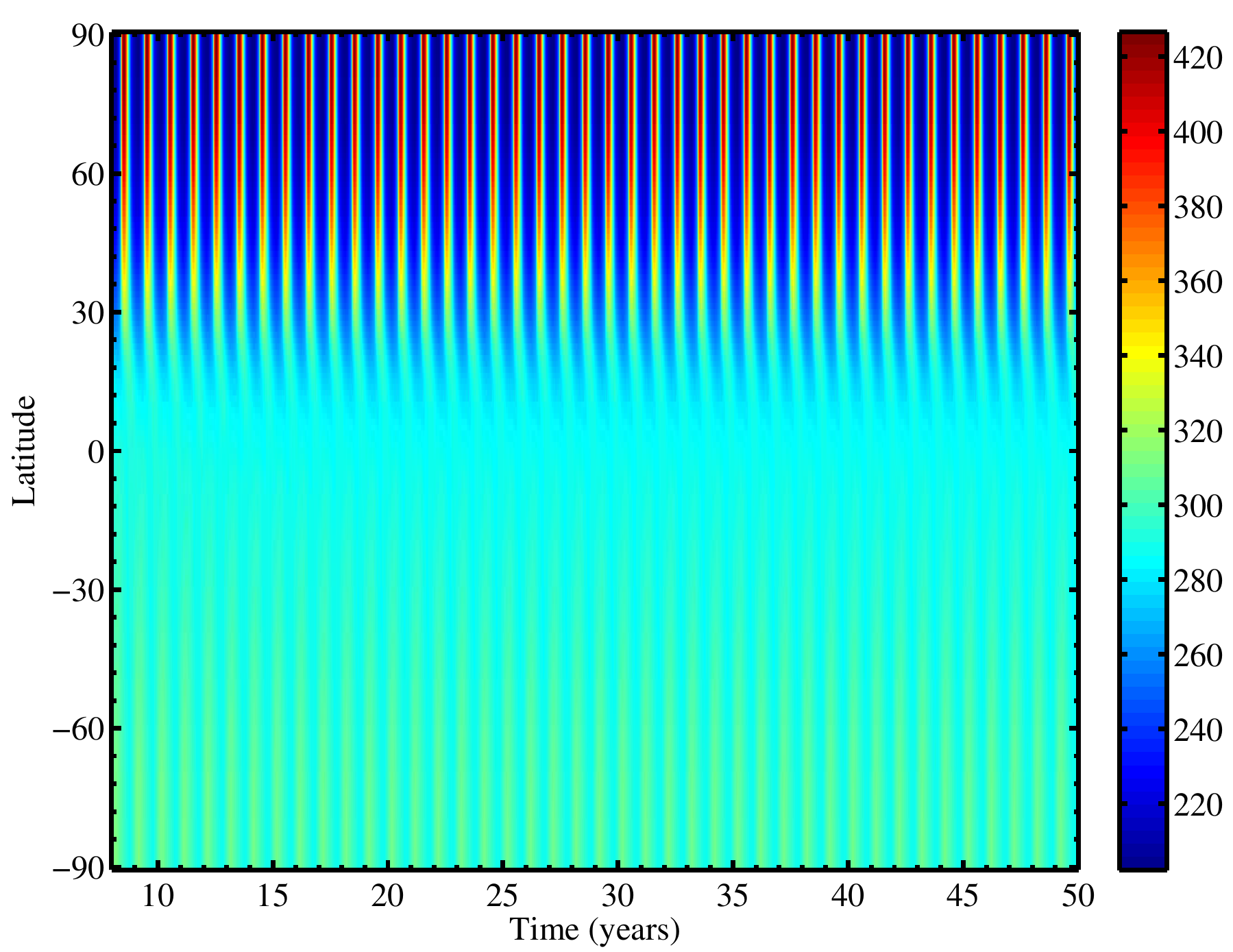}\\
\plottwo
{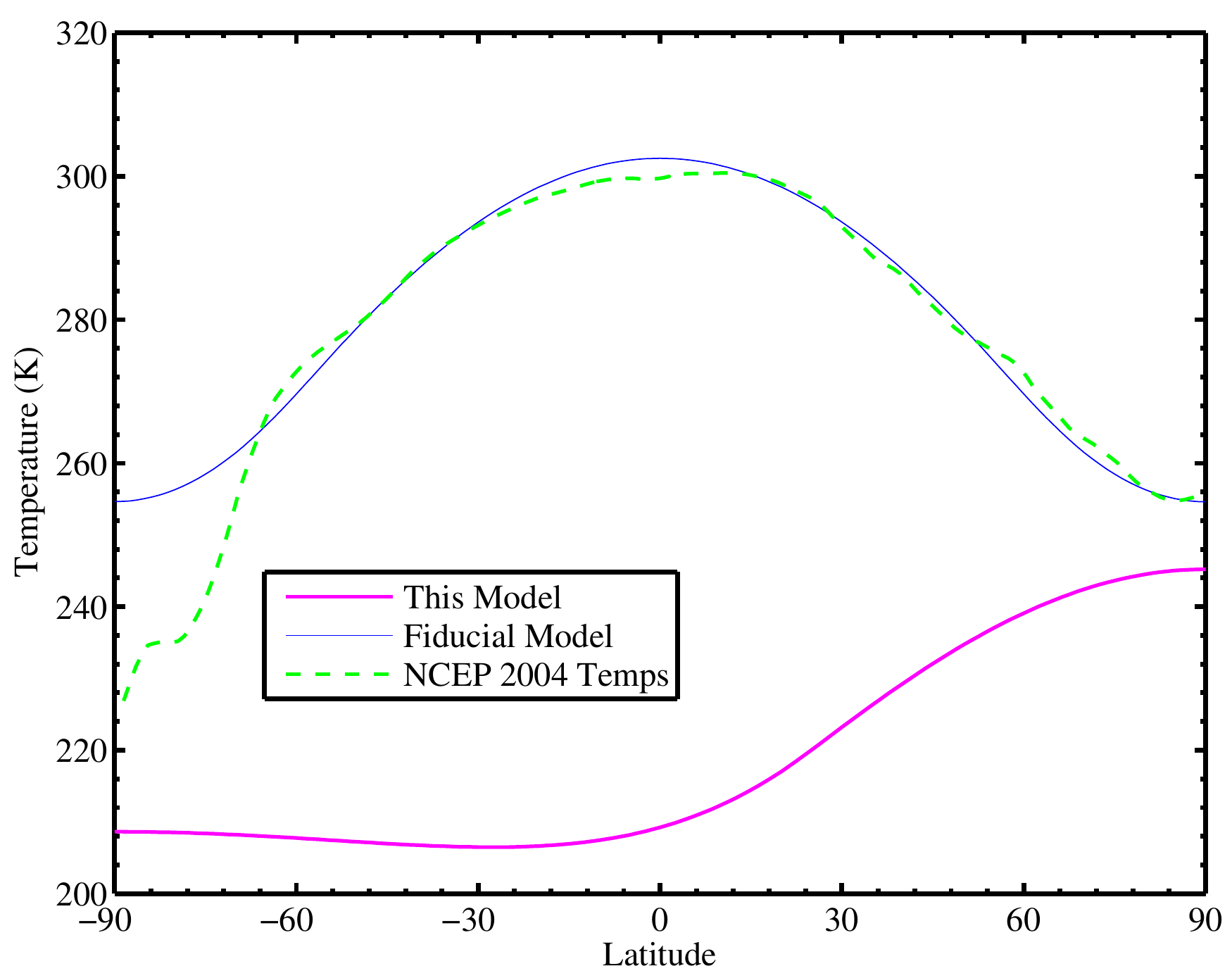}
{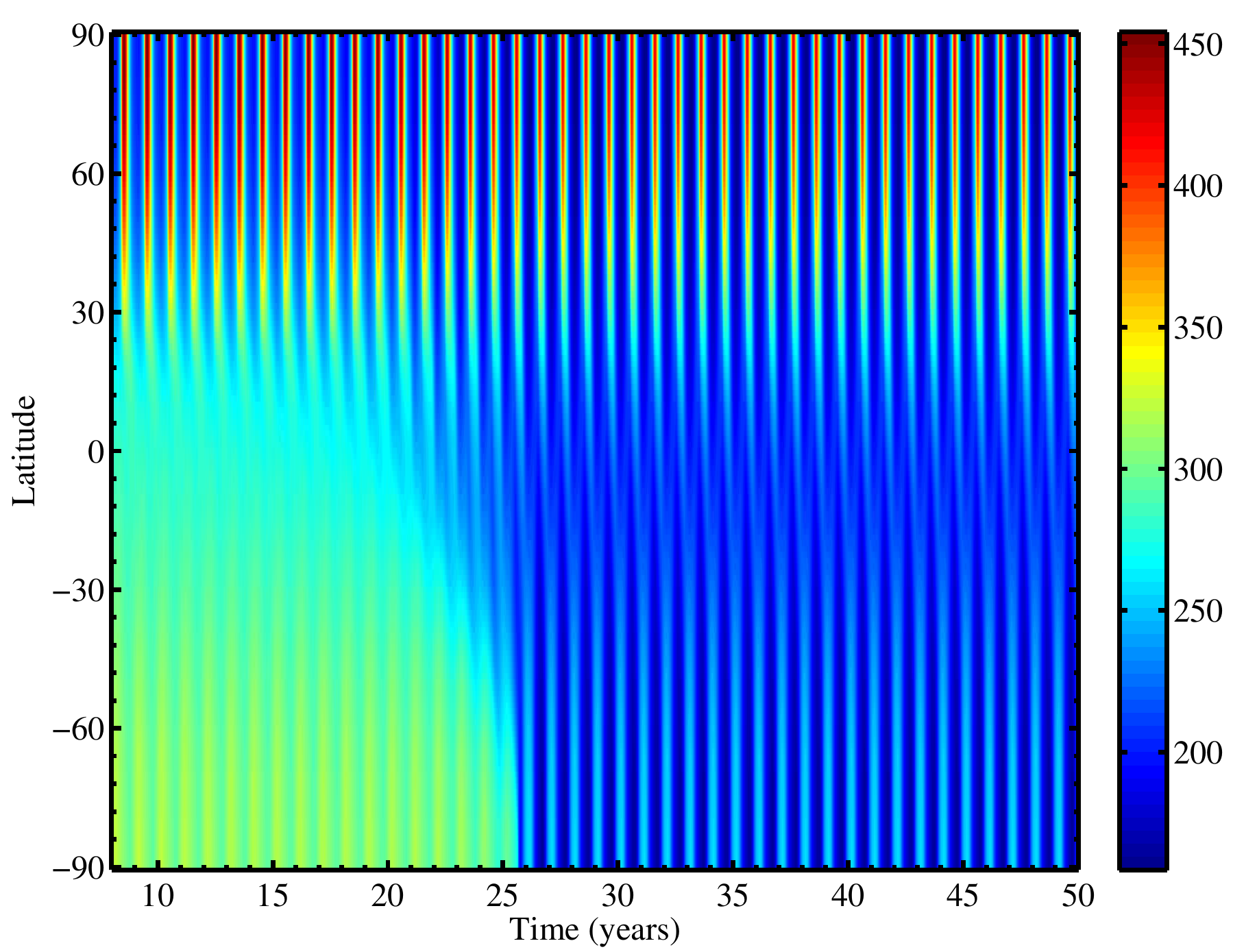}\\
\caption[Annually averaged and space-time plot of temperatures on
models with a North Polar continent that takes up 30\% of surface area
at 1~AU, cooling/heating = $I_2,A_2$, obliquity = $23.5\degr$,
$60\degr$, $90\degr$.]{\footnotesize Annually averaged and detailed
time-dependent temperature profiles in models with a North Polar
continent that takes up 30\% of surface area (the other 70\% is ocean)
at 1~AU orbital distance, for ($I_2,A_2$) cooling-albedo functions and
obliquities of $23.5\degr$, $60\degr$ and $90\degr$. {\it Left:} The
magenta curve shows the annually averaged temperature profile for
models with a North Polar continent that extends down to $\sim
24\degr$ North latitude.  The solid blue and dashed green curves are
for reference -- blue: the fiducial model (identical to this one,
except for 70\% ocean uniformly distributed in every latitude band);
dashed green: the Earth's actual temperature profile, as measured by
NCEP/NCAR in 2004 \citep{kistler_et_al1999, kalnay_et_al1996}.  {\it
Right:} Detailed temperature evolution (latitude vs. time) in the
polar continent models from years 8 through 50. {\it Top Row:
$23.5\degr$.}  {\it Middle Row: $60\degr$.}  {\it Bottom Row:
$90\degr$.} Annually averaged profiles (left) miss much of the
seasonal variations shown by the detailed profiles (right). The
$90\degr$ obliquity model experiences an asymmetric, partial snowball
transition.

}
\label{obl_fig:set31_23p5_60_90}
\end{figure}

Figure~\ref{obl_fig:set31_23p5_60_90} displays the tremendous swings
of temperature that can occur over latitude bands that lack ocean, and
also indicates that annually averaged calculations can miss a lot of
information about instantaneous conditions on oblique planets.  This
figure contrasts the annually averaged temperature with the detailed
temperature evolution on a model planet with a North Polar continent
that is 30\% of the total surface area, at an orbital distance of
1~AU.  This model uses ($I_2,A_2$) cooling--albedo functions and
results are shown for obliquities $23.5\degr$, $60\degr$, and
$90\degr$.  At all three obliquities, the left column -- the annually
averaged temperature profile -- provides an impoverished view of the
actual climatic conditions.  Looking at just the left panels: the
$23.5\degr$ obliquity model appears slightly asymmetrical in
temperature distribution, with the continental North Pole 8~K warmer
than the oceanic South Pole; the $60\degr$ obliquity model appears
cooler at the continental pole; and the $90\degr$ obliquity model
again appears warmer at the continental pole, but appears frozen over
the whole globe.  In truth, all three models reach significantly
higher temperatures at the continental pole during its summer than at
the other pole.  At both $60\degr$ and $90\degr$ obliquity, North Pole
summer temperatures exceed 410~K, as the Sun shines nearly straight
down on the pole for months.  We note that an important limitation of
our models is apparent in this figure. Although we may not have much
intuition for what the polar summers should be like on high obliquity
planets, it is surprising to obtain summer polar continent
temperatures in excess of 310~K in the $23.5\degr$ obliquity model.
Indeed, Antarctica -- Earth's continental pole -- is significantly
{\it colder} than the non-continental pole, and for the most part
neither pole ever reaches temperatures above freezing.  Accounting for
Antarctica in our model framework would be non-trivial, as it may
require initial conditions differing from uniformly ice-free and/or
improved treatments of ice-related surface processes.

\begin{figure}[p]
\plotone
{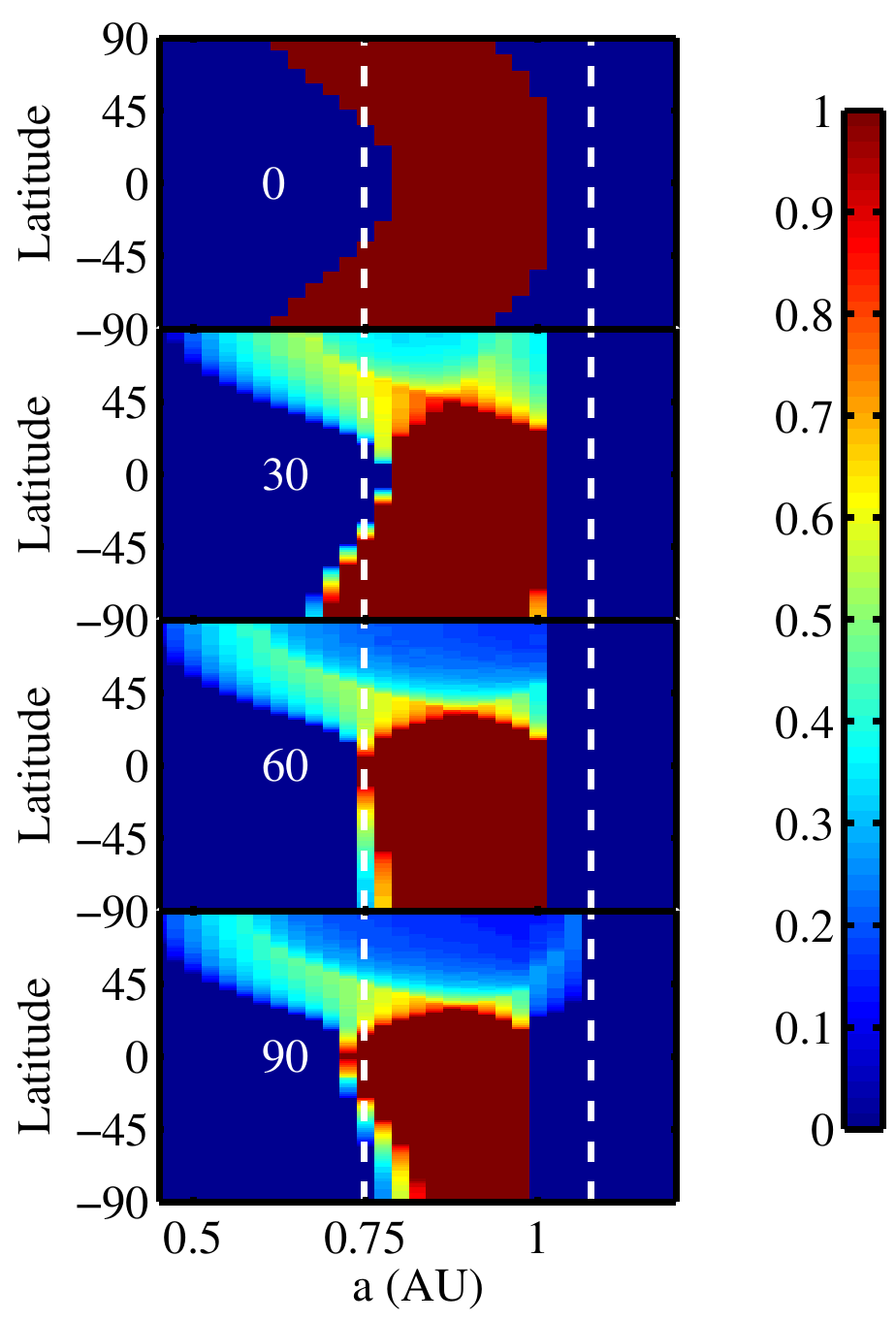}
\caption[Model temporal habitability fraction under different
obliquities, North Polar continent covering 30\% of surface.]{Temporal
habitability fraction at different obliquities for a model planet with
a North Polar continent covering 30\% of its surface. Obliquity varies
from $0\degr$ to $90\degr$ from top to bottom. The latitudinal heat
diffusion coefficient is kept to its fiducial value.  The notation is
similar to Figure~\ref{obl_fig:obliq_rotI2A2I3A3}: colors indicate the
fraction of the year spent by that region in the habitable temperature
range (273~K - 373~K). Habitability has a strong regional and seasonal
character on such a planet, when oblique.}
\label{obl_fig:set26_obliq}
\end{figure}

Figure~\ref{obl_fig:set26_obliq} presents plots of the temporal
habitability fraction for the same model planet as above, at
obliquities $0\degr$, $30\degr$, $60\degr$, $90\degr$.
Figure~\ref{obl_fig:set31_23p5_60_90} illustrated how the presence of
land at the North Pole causes tremendous swings in temperature there;
Fig.~\ref{obl_fig:set26_obliq} confirms that, for nonzero obliquity,
this is indeed the case throughout the orbital extent of the habitable
zone.  These large seasonal variations lead to exotic shapes in plots
of temporal habitability.  Compared with uniformly ocean-dominated
worlds, much more of the parameter space is partially habitable at
each obliquity (except $0\degr$), neither 0\% nor 100\% of the year,
but somewhere in between.

\subsubsection{Desert Worlds}
\label{obl_sssec:desert}

\begin{figure}[p]
\plotone
{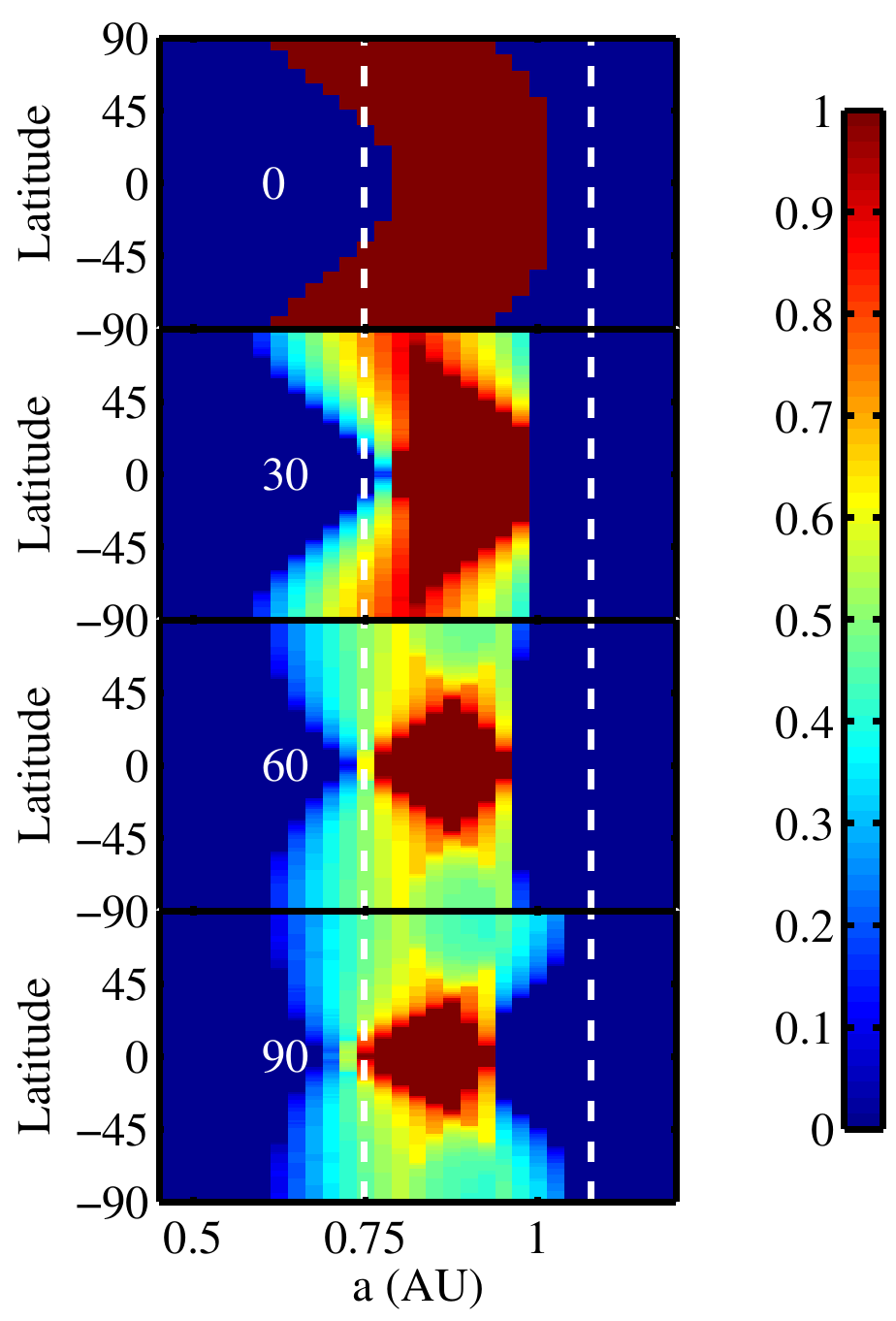}
\caption[Model temporal habitability fraction under different
obliquities, 10\% ocean uniformly distributed.]{Temporal habitability
fraction at different obliquities for a model planet with 10\% ocean
uniformly distributed.  Otherwise similar to
Figure~\ref{obl_fig:set26_obliq}. Habitability has a strong regional
and seasonal character on such a planet, when oblique.}
\label{obl_fig:set25_obliq}
\end{figure}

\begin{figure}[p]
\plotone
{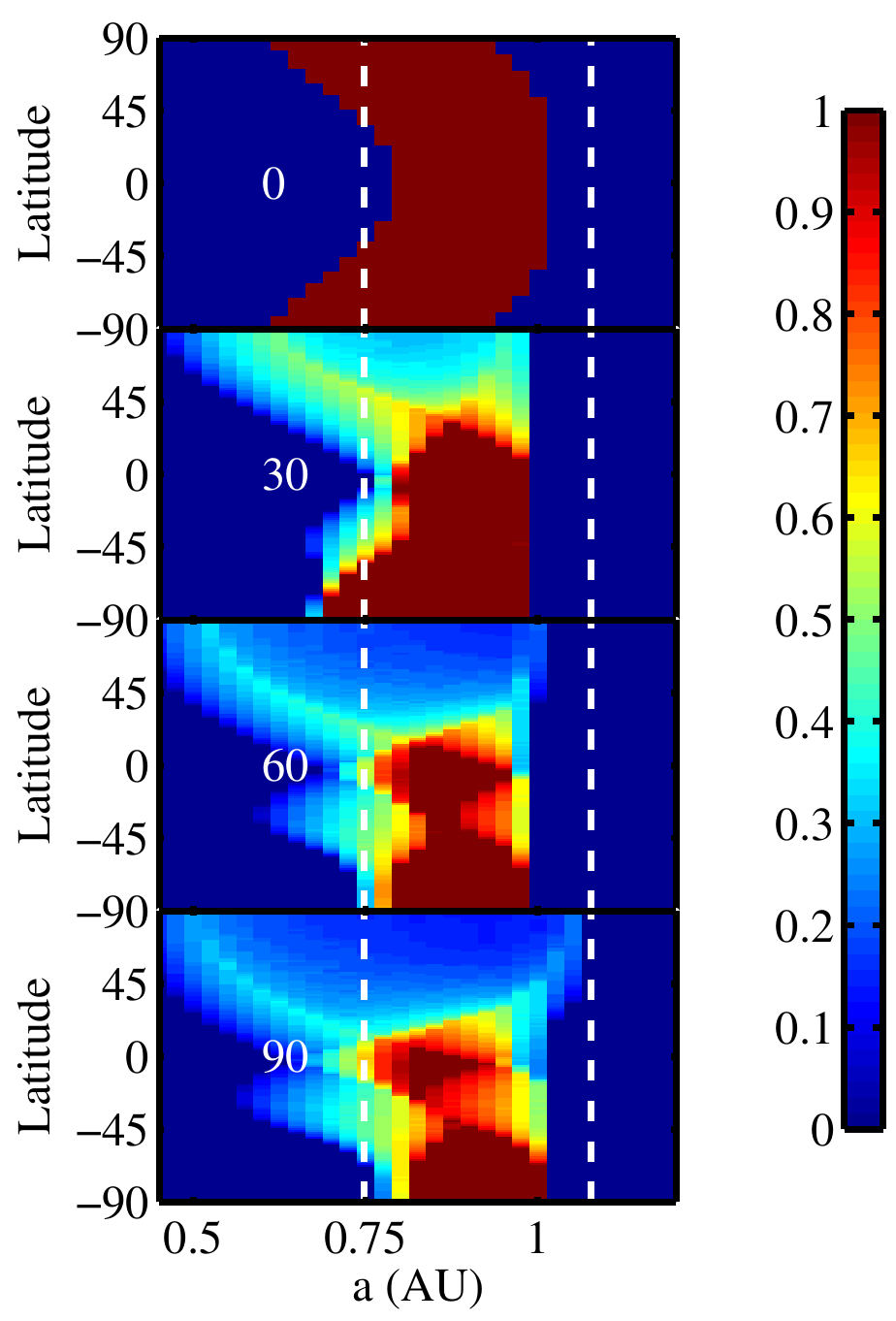}
\caption[Model temporal habitability fraction under different
obliquities, North Polar continent covering 90\% of surface (South
Polar ocean).]{Temporal habitability fraction at different obliquities
for a model planet with a North Polar continent covering 90\% of its
surface (i.e., a localized South Polar ocean).  Otherwise similar to
Figure~\ref{obl_fig:set26_obliq}. Habitability has a strong regional
and seasonal character on such a planet, when oblique.}
\label{obl_fig:set28_obliq}
\end{figure}

We now consider two model planets with just 10\% ocean fraction.  We
examine the cases of a uniformly distributed ocean (10\% in every
latitude band) and an ocean localized at the South Pole (extending
northward to $\sim 53\degr$ South latitude).
Figure~\ref{obl_fig:set25_obliq} presents the temporal habitability
for the uniform desert world, while Figure~\ref{obl_fig:set28_obliq}
presents the analogous plot for the desert world with a South Polar
ocean.  As before, some regions of these model planets swing from
freezing to boiling temperatures over the course of the year.  This is
responsible for the butterfly shape of the temporal habitability plots
in the $60\degr$ and $90\degr$ obliquity cases shown in
Fig.~\ref{obl_fig:set25_obliq}: at $a\sim 0.9\rm~AU$, the poles are
habitable for a smaller fraction of the year than more equatorial
regions at that orbital distance, or than the poles for closer and
more distant orbits. The pattern of habitability in
Fig.~\ref{obl_fig:set28_obliq}, on the other hand, shows how strongly
assymmetric the climate can be on a desert world with a polar ocean.

These models, and those presented in \S~\ref{obl_sssec:nonuniform},
suggest that at extreme obliquity the inner edge of the zone of
regionally and seasonally habitable climates can be extended
dramatically inward, while the outer boundary can only be extended
mildly outward.  Several important caveats should accompany this
observation, however.  Assuming an infrared cooling function that is
constant with orbital radius probably leads to a flawed treatment at
both the high and low insolation limits of these models.  At the inner
edge of the habitable zone, large increases in atmospheric water
content can cause a reduction in the cooling efficiency, leading to a
runaway greenhouse effect. An eventual catastrophic water loss can
result in a Venus-like outcome, as described by
\citet{kasting_et_al1993} and references therein (although this type
of outcome might be mitigated by reduced heating from increased
cloud-albedo, as mentioned in SMS08).  At the outer edge, on long
timescales, a reduced efficiency of the carbonate-silicate weathering
cycle is likely to lead to a significant increase in the partial
pressure of atmospheric CO$_2$ \citep{kasting_et_al1993}, which could
extend the habitable zone from $\sim 1\rm~AU$, as in our models, to
$\sim 1.4\rm~AU$ or beyond in some cases.  We discuss in greater
detail the issue of varying atmospheric CO$_2$ content with orbital
distance in \S~\ref{obl_ssec:IWK97}.

Notwithstanding these various complications, for plausible cooling
functions, the low thermal inertia of atmosphere over land might lead
to severe polar climates on highly oblique planets.  What are we to
make of partial ``habitability'' by our criterion in the case of a
region of a planet that actually boils and freezes every year?  There
are some microbes on Earth that can reproduce at freezing
temperatures, and others that can reproduce at boiling temperatures
(see SMS08 and references therein), although none of which we are
aware that can do both.  If part of a planet regularly swings through
these wild extremes of climate, is it appropriate to call it
habitable?  More relevant from the perspective of formulating a
testable scientific hypothesis, could such a planet support enough
life to produce sufficient levels of biosignatures that its life could
be detected from Earth?  This is an open question, but it is
worthwhile to keep in mind that microbes on Earth appear to be as
hardy as they need to be: nearly everywhere that biologists have
searched, they have found some microbes thriving.  A perhaps equally
significant result from the perspective of habitability is that the
reduced thermal inertia of these models appears to render them
somewhat less susceptible to global snowball events, especially at
high obliquities (e.g., compare
Figs.~\ref{obl_fig:set26_obliq}-\ref{obl_fig:set28_obliq} with the
middle column of Fig.~\ref{obl_fig:obliq_rotI2A2I3A3}).

\subsection{Modeling the Far Reaches of the Habitable Zone}
\label{obl_ssec:IWK97}

\citet{walker_et_al1981} propose that a planet's temperature is
regulated on long timescales by a feedback mechanism involving
weathering of silicate rocks through carbonic acid from CO$_2$
disolved in water.  They argue that since the rate of weathering (and
hence of removal of CO$_2$ from the atmosphere) increases with
temperature, this process is an important negative feedback on climate
that acts to keep temperatures near the freezing point of water (see
the recent study by \citealt{zeebe+caldeira2008} confirming the
operation of this cycle).  \citet{kasting_et_al1993} point out that
this negative feedback can significantly offset the extreme
sensitivity of climate to changes in orbital distance away from 1~AU
seen in models such as those of \citet{hart1979} and the models
presented in SMS08 and thus far in this paper.  These models are
sensitive because they contain a significant positive feedback of the
Earth-climate -- the ice-albedo feedback whereby at lower temperatures
the absorbed insolation is dramatically reduced because of the high
albedo of ice.  However, these models also ignore the long-term,
negative (counterbalancing) feedback of the carbonate-silicate
weathering cycle \citep{kasting_et_al1993}.

In order to probe the combined influence on climate of rotation rate
and obliquity in the context of the expected CO$_2$-rich atmosphere
that a pseudo Earth would have at 1.4~AU, we switch to the infrared
cooling function used by WK97 ($I_{\rm WK97}$ in
Table~\ref{obl_tab:one}), with CO$_2$ partial pressure ($p$CO$_2$) set
to 1~bar and 2~bars. For simplicity, we maintain the same albedo
function $A_2$ as before and adopt the same linear dependence of the
latitudinal heat diffusion coefficient with total atmospheric pressure
as WK97 ($D/D_{\rm fid} \propto P_{\rm tot}$).  We find that at both
1- and 2-bar levels of CO$_2$, model planets maintain globally
temperate conditions at all obliquities for both $D=9D_{\rm fid}$ and
$D=D_{\rm fid}$ -- corresponding to slow and Earth-like rotation.
Interestingly, however, at reduced transport efficiency, corresponding
to fast planetary rotation, we find that these CO$_2$-rich models are
susceptible to global glaciation events.

\begin{figure}[p]
\plottwo
{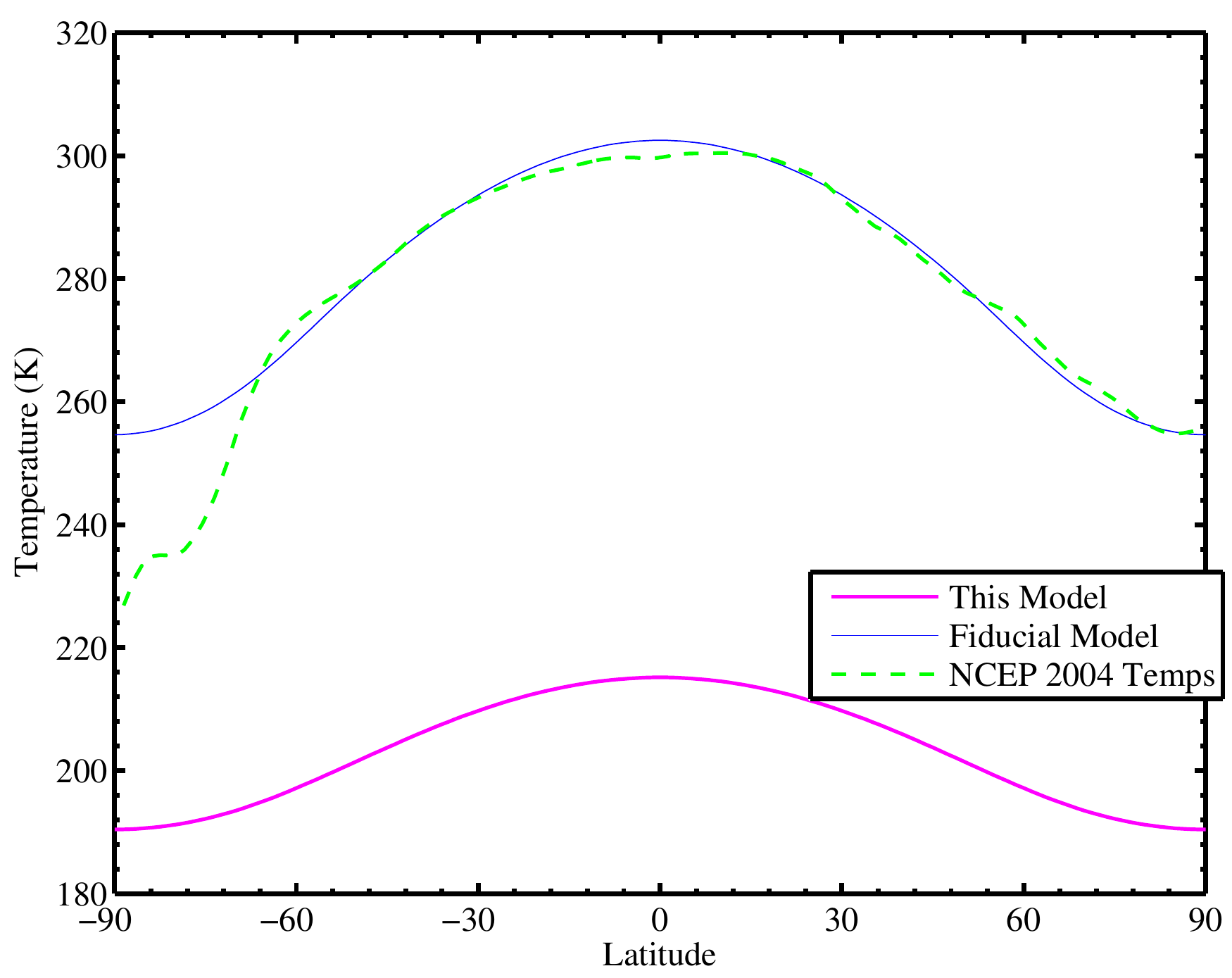}
{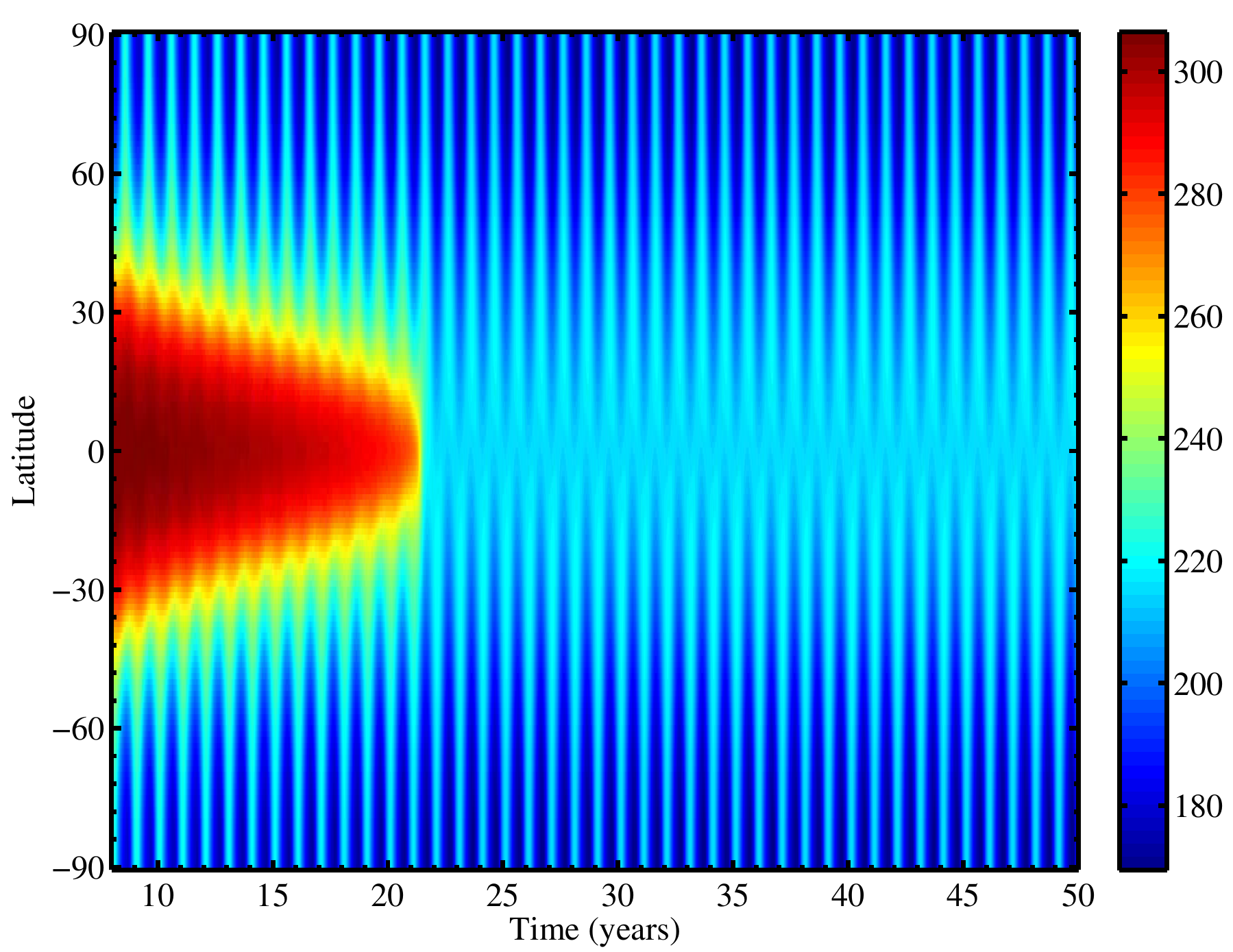}\\
\plottwo
{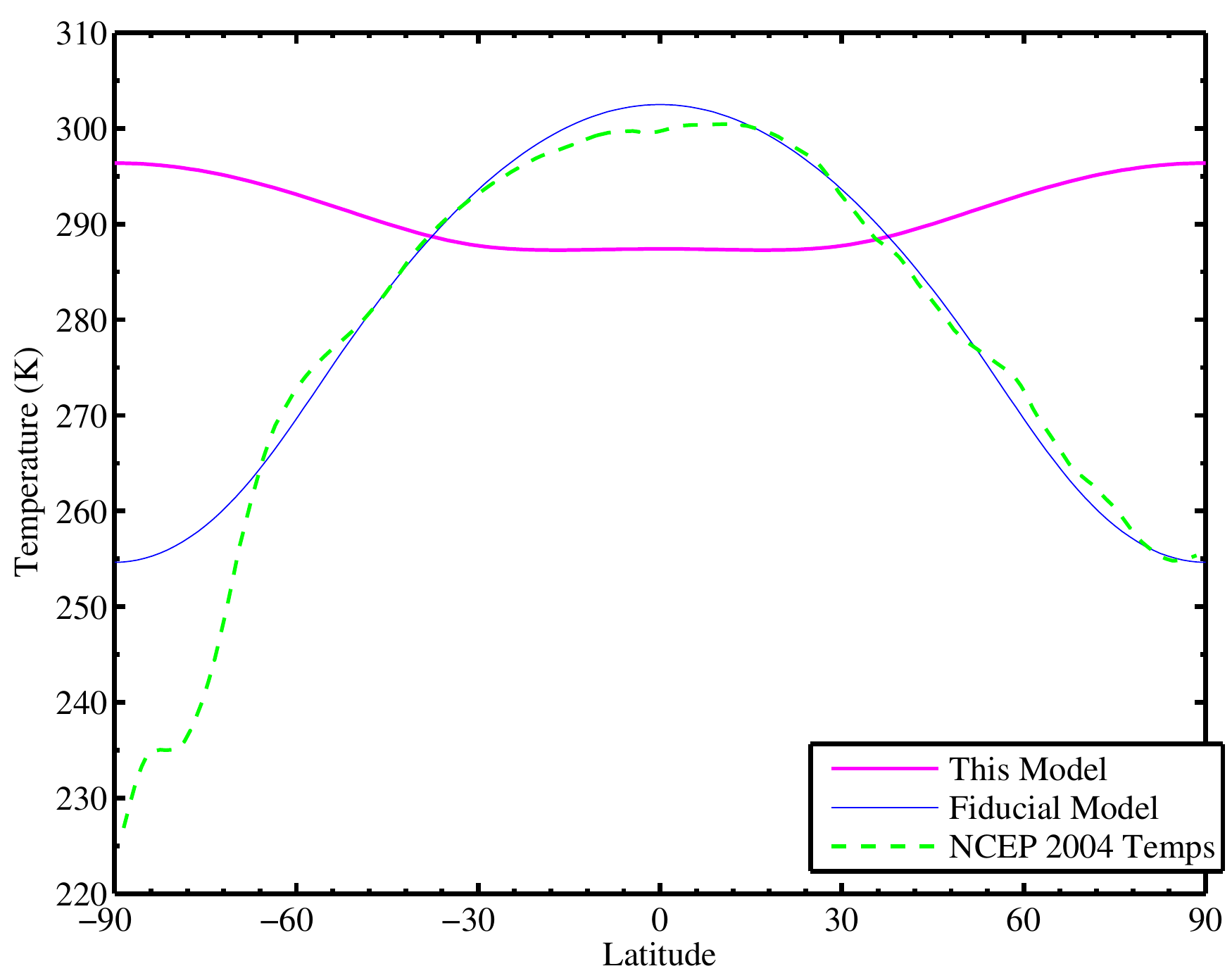}
{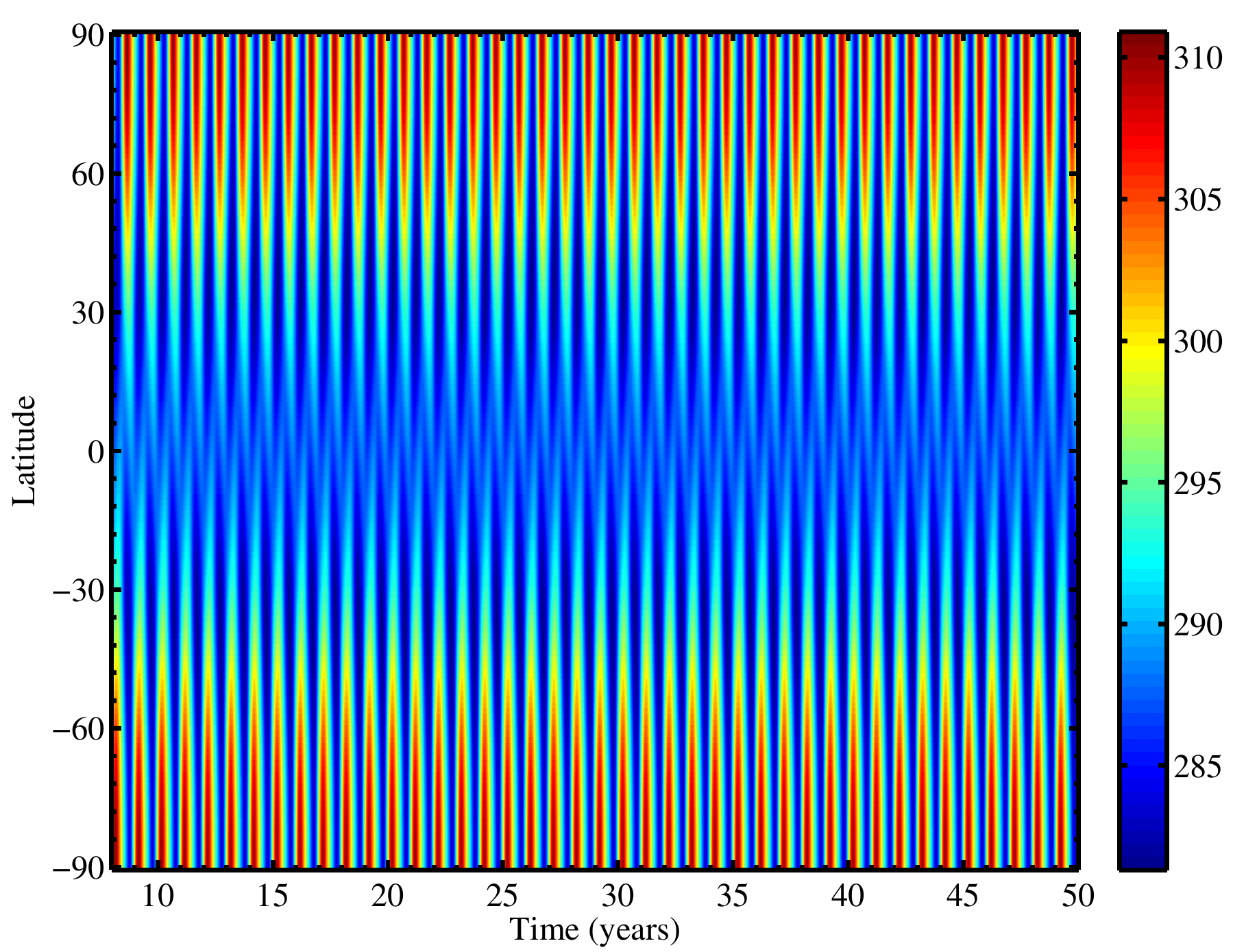}\\
\plottwo
{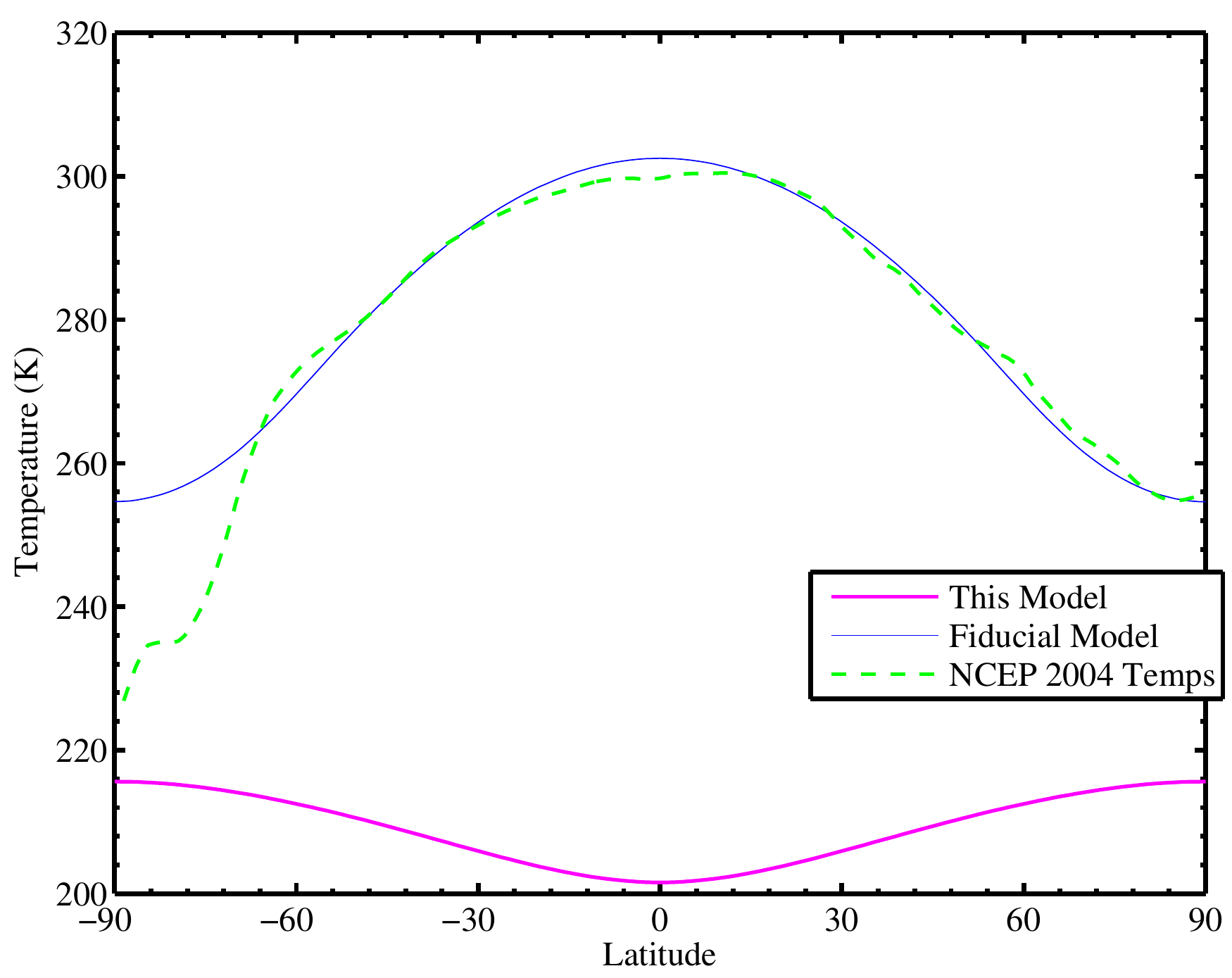}
{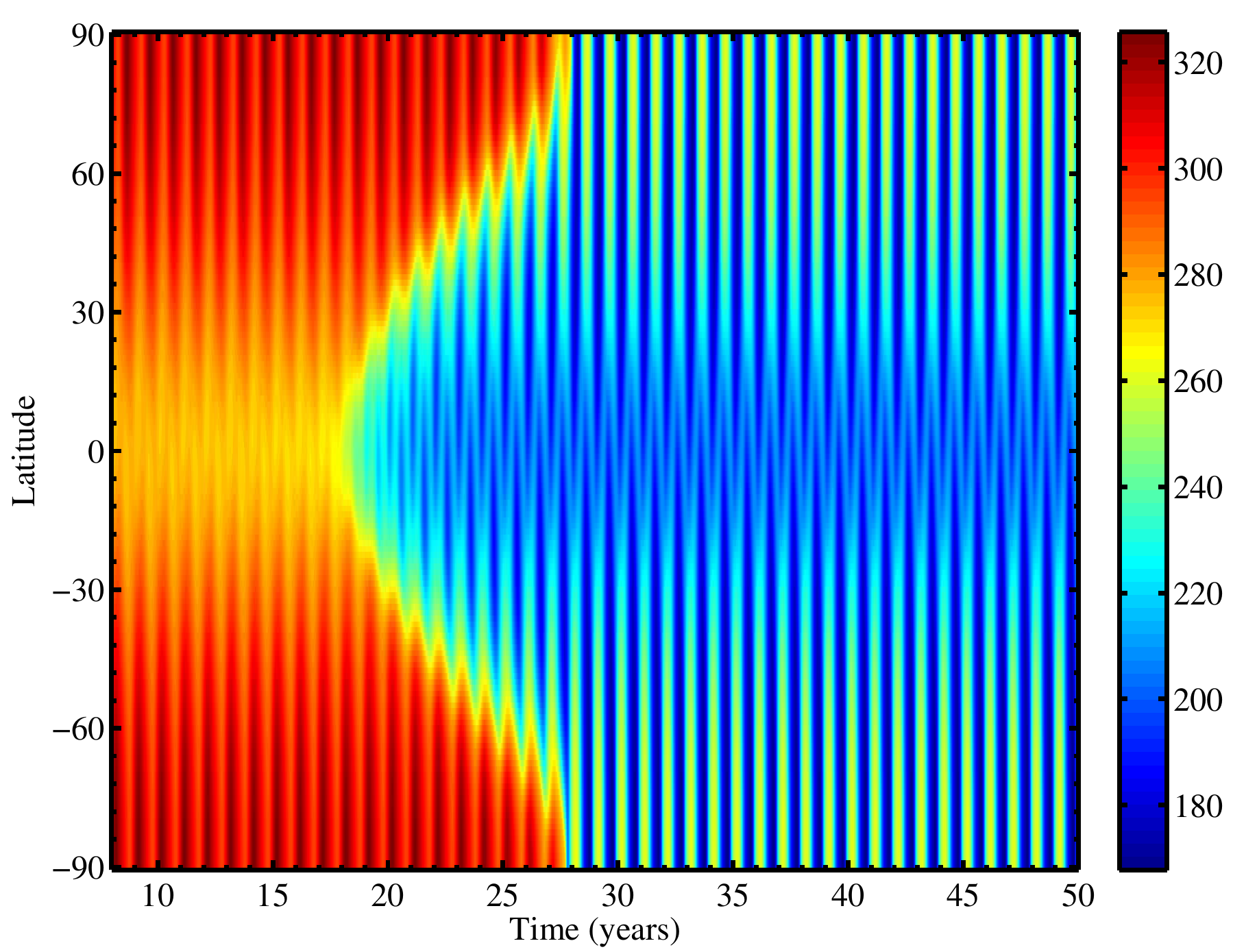}
\caption[Annually averaged and space-time plot of temperatures on fast
spinning world at 1.4~AU, WK97 cooling function, $p\rm CO_2 = 1~bars$,
obliquity = $23.5\degr$, $60\degr$, $90\degr$.]{Annually averaged and
detailed time-dependent temperatures on a fast spinning model planet
at 1.4~AU, using the infrared cooling function of WK97 for a
CO$_2$-rich atmosphere with $p\rm CO_2 = 1~bar$. From top to bottom,
results are shown for an obliquity of $23.5\degr$, $60\degr$ and
$90\degr$.
The notation and the models are the same as in
Figure~\ref{obl_fig:set31_23p5_60_90}, except for the cooling function
($I_{\rm WK97}$ with $p\rm CO_2 = 1~bar$) and a latitudinal heat
diffusion coefficient adjusted for fast spin and a massive
atmosphere. In the two globally-frozen models (top and bottom),
surface temperatures that would result in seasonal CO$_2$ atmospheric
collapse ($T \lsim 195$~K) are reached over a fair fraction of the
planet's surface area.}
\label{obl_fig:set45_23p5_60_90}
\end{figure}

Figure~\ref{obl_fig:set45_23p5_60_90} shows the global average
temperature and the climate evolution for fast-spinning model planets
at 1.4~AU with 1~bar atmospheric CO$_2$, ($I_{\rm WK97},A_2$)
cooling-albedo functions and a latitudinal heat diffusion coefficient
scaled up with total pressure but reduced by a factor 9 to account for
rapid rotation.  At $23.5\degr$ obliquity, the cold temperatures at
the poles drag the model into a snowball state, while at $90\degr$
obliquity, it is the cold equator that drags the model planet to the
same fate. At $60\degr$ obliquity, however, no part of the planet
receives consistently low enough insolation to trigger global
glaciation.  The snowball effect seen in the $23.5\degr$ and $90\degr$
obliquity models might be particularly dramatic because of the
possibility of partial atmospheric collapse, on a much shorter
timescale than volcanism can replenish CO$_2$.  At 1~bar, the freezing
point of CO$_2$ is $\sim 195 \rm~K$.  Both the $23.5\degr$ and
$90\degr$ obliquity models reach temperatures below this threshold
over large enough regions of their surfaces\footnote{The very large
seasonal variations of temperature on these snowball planets result
from the moderate heat capacity of the atmosphere over ice (see, e.g.,
SMS08; WK97).} that significant amounts of the atmospheric CO$_2$
might condense out as dry ice, thereby reducing the atmospheric
greenhouse effect.  The risk of atmospheric collapse would be somewhat
lessened by the release of latent heat from CO$_2$ condensation, which
would tend to prevent too much CO$_2$ from freezing out during any
winter. Partial collapse, on the other hand, would reduce surface
pressure and thus the efficiency of atmospheric heat
transport. Realistically treating these possibilities is beyond the
scope of the present paper as it would require incorporating a latent
heat term in the energy balance equation and accounting for the
surface CO$_2$ mass budget, as \citet{nakamura+tajika2002} do in their
Mars EBM.

While model runs with 2~bars of CO$_2$ (not shown) indicate that a
pseudo-Earth with such a massive atmosphere at 1.4~AU would be
unlikely to suffer glaciation or partial atmospheric collapse at any
obliquity, it is worth noting that on a planet with a rate of volcanic
greenhouse gas output not much higher than on Earth, it may be
difficult to build up a thick CO$_2$ atmosphere (e.g., 2 bars) under
conditions such that CO$_2$ condenses out at lower pressure (like in
our model with 1 bar of CO$_2$).  In situations in which atmospheric
CO$_2$ concentrations can change appreciably on a yearly timescale,
our model is not self-consistent.  Further analysis is therefore
needed to determine the full extent and consequences of such
atmospheric condensation events.  Clearly, these various issues are
important subjects for future studies since they indicate that
snowball transitions could in principle limit the habitability of some
terrestrial planets by interfering with the negative feedback of their
carbonate-silicate weathering cycles.

\subsection{Validity of Global Radiative Balance}
\label{obl_sssec:global_balance}

Calculations of habitable zones have often assumed global radiative
balance conditions.  Although these calculations by definition cannot
account for the regional character of habitability, one might hope
that they still provide a decent proxy for the global average
conditions.  Indeed, as seen in
Fig.~\ref{obl_fig:set30_heat_cool_vs_time}, the Earth itself is within
$\sim 5\%$ of radiative equilibrium throughout the year.  Accordingly,
global radiative balance models have provided a very useful starting
point for considerations of how the habitability of an Earth-like
planet depends on its orbital radius.

\begin{figure}[p]
\plotone
{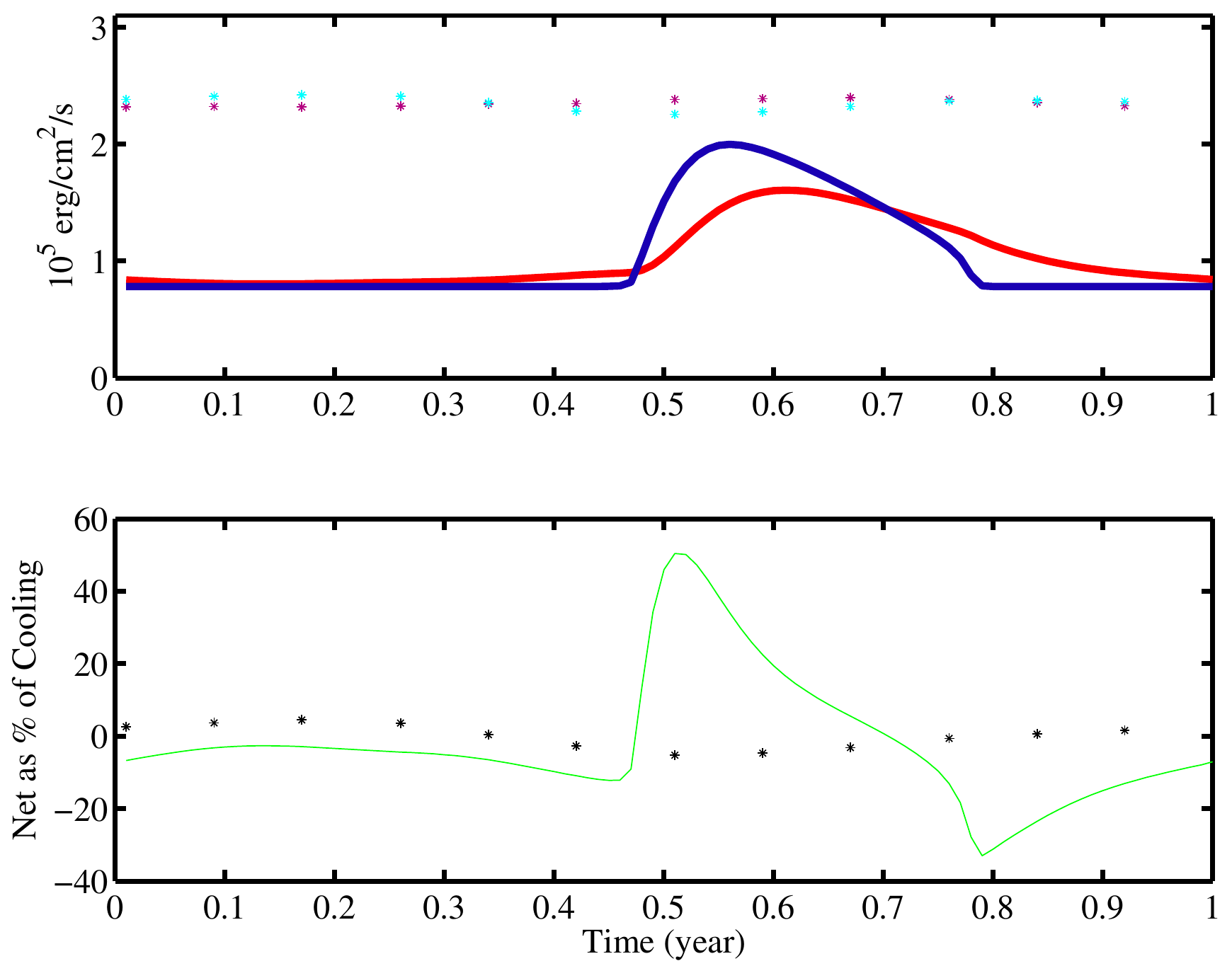}
\caption[Global average cooling, heating, and net radiative flux, as
functions of time in model with North Polar continent at $60\degr$ and
$90\degr$ obliquity.]{Global average cooling, heating, and net
(heating $-$ cooling) radiative flux, as functions of time, for a
model planet with a North Polar continent covering $30 \%$ of its
surface, at $90\degr$ obliquity.
Notation is similar to
Figure~\ref{obl_fig:set30_heat_cool_vs_time}. This planet experiences
large deviations from global radiative balance.}
\label{obl_fig:set31_heat_cool_vs_time}
\end{figure}

Figure~\ref{obl_fig:set31_heat_cool_vs_time} presents globally
averaged cooling, heating, and net radiative fluxes in a model with a
North Polar continent that covers 30\% of the planet's surface, with
$90\degr$ obliquity.  In contrast to the Earth, which remains within
$5\%$ of global radiative balance, this model planet can be nearly
60\% out of global radiative balance.  While it has been recognized
that planets on highly eccentric orbits experience forcings that are
significantly different from annually and globally averaged
conditions, the results in Fig.~\ref{obl_fig:set31_heat_cool_vs_time}
illustrate how, even on circular orbits, planets can experience
conditions that are far from radiative equilibrium.  This further
underscores the importance of regional, time-dependent climate models
for addressing the habitability of extrasolar terrestrial planets.

\section{Conclusions}
\label{obl_sec:disc+conc}

We have presented a series of energy balance models to address the
variety of climatic conditions that might exist on oblique terrestrial
planets with circular orbits.  We considered dynamic climate forcings
and responses determined by several planetary attributes a priori
unknown for extrasolar planets, including obliquity, rotation rate,
distribution of land/ocean coverage, and the detailed nature of the
radiative cooling and heating functions.  We find that planets with
small ocean fractions or polar continents can experience very severe
seasonal climatic variations, but that these planets also might
maintain seasonally and regionally habitable conditions over a larger
range of orbital radii than more Earth-like planets.  Climates on high
obliquity planets with nonuniform distributions of land and ocean can
be far from global radiative balance, as compared to the Earth. Our
results provide indications that the modeled climates are somewhat
less prone to dynamical snowball transitions at high obliquity.  Fast
rotating Earth-like planets may fall victim to global glaciation
events at closer orbital radii than slower rotating planets. This is
also the case for planets with massive CO$_2$ atmospheres, which are
expected to be found in the outer orbital range of habitable zones.
Snowball transitions could be particularly significant for such
planets since partial collapse of their CO$_2$-rich atmospheres may
occur and possibly interfere with the thermostatic effect of their
carbonate-silicate weathering cycle, thus affecting their long-term
habitability.

\section*{Acknowledgments}
We acknowledge helpful conversations with James Cho, Michael Allison,
Anthony Del Genio, and Scott Gaudi.  We thank Diana Spiegel for help
with ERBE data.  We acknowledge many useful comments by an anonymous
referee. CS acknowledges the funding support of the Columbia
Astrobiology Center through Columbia University's Initiatives in
Science and Engineering, and a NASA Astrobiology: Exobiology and
Evolutionary Biology; and Planetary Protection Research grant, \#
NNG05GO79G.


\clearpage

\begin{table}[p]
\begin{center}
\caption[Atmospheric Models]{~~Atmospheric Models.}
\vspace{0.2in}
\begin{tabular}{cll}
  \tableline
  \tableline
  Model             & IR Cooling Function      & Albedo Function\\[0.1in]
  \tableline
2\tablenotemark{a}  &  $I_2[T]  =  \frac{\sigma T^4}{1 + (3/4)\tau_{\rm IR}[T]}$   &  $A_2[T] =  0.525 - 0.245 \tanh[\frac{(T-268{\rm K})}{5{\rm K}}]$   \\
3\tablenotemark{b}  &   $I_3[T]  =  A + B T$   &  $A_3[T] =  0.475 - 0.225 \tanh[\frac{(T-268{\rm K})}{5{\rm K}}]$  \\
4\tablenotemark{c}  &   $I_{\rm WK97}[T,p\rm{CO_2}]$   &  $A_2[T]$  \\
\label{obl_tab:one}
\end{tabular}
\vspace{-0.4cm}
\tablenotetext{a}{\,Model with $T$--dependent optical thickness: $\tau_{\rm IR}[T] = 0.79(T/273{\rm K})^3$.}
\tablenotetext{b}{\,Linearized model: $A = 2.033\times 10^5{\rm~erg~cm^{-2}~s^{-1}}$, $B = 2.094\times 10^3 {\rm~erg~cm^{-2}~s^{-1}~K^{-1}}$.}
\tablenotetext{c}{\,WK97 cooling function (with $A_2$ albedo). The detailed functional form is presented in the appendix of WK97.}
\tablecomments{$\sigma$ is the Stefan-Boltzmann constant.}
\end{center}
\end{table}

\clearpage

\clearpage
\bibliography{biblio.bib}

\end{document}